\documentclass[usenatbib,usegraphicx]{mn2e}
\def\mnras{MNRAS}
\def\apj{Ap.J}
\def\dt{\tilde{\eta}_{\rm HI}}
\def\u{{\bf U}}
\def\th{\vec{\theta}}
\def\x{{\bf x}}
\def\HI{{\rm HI}}
\def\H{{\rm H}}
\def\mK{\rm mK}
\def\d{\eta_{\HI}}

\def\n{{\bf n}}
\def\Tc{T_\gamma}
\def\Tb{T_b}
\def\Ts{T_s}
\def\Tg{T_g}
\def\nue{\nu_e}
\def\Bon{B_{10}}
\def\Bo{B_{01}}
\def\A10{A_{10}}
\def\hp{h_p}
\def\kb{k_B}
\def\rn{r_{\nu}}
\def\rnp{r_{\nu}^{'}}
\def\nH{n_{HI}}
\def\k{{\bf k}}
\def\kp{k_\parallel}
\def\kpr{{\bf k}_\perp}
\usepackage{epsfig}
\usepackage{graphics}
\begin{document}
\title[Using visibility correlations to probe HI distribution]{ On
  using visibility correlations to probe the HI distribution 
  from the dark ages to the present epoch  I: Formalism and the
  expected signal}
\author[S. Bharadwaj and S. S. Ali]{Somnath Bharadwaj\thanks{Email:
    somnathb@iitkgp.ac.in} and SK. Saiyad Ali\thanks{Email:
    saiyad@cts.iitkgp.ernet.in}
  \\ Department of Physics and Meteorology\\ and
  \\ Centre for Theoretical Studies \\ IIT Kharagpur \\ Pin: 721 302 ,
  India }
\maketitle
\begin{abstract}
Redshifted 21 cm   radiation originating from the cosmological
distribution  of neutral hydrogen (HI)  appears as a background
radiation in low frequency radio observations. The  angular and
frequency domain fluctuations in this radiation  carry  information
about cosmological structure formation. We propose that correlations
between visibilities measured at different baselines and frequencies
in radio-interferometric observations be used to quantify the
statistical properties of these fluctuations.  This has an inherent
advantage over other statistical estimators in that it deals directly
with the visibilities which are the primary quantities measured in
radio-interferometric observations. Also, the visibility correlation
has a very simple relation with the power spectrum. We present estimates
of the expected signal for  nearly the  entire post-recombination era,
from the dark ages to the present epoch. The epoch of reionization, 
where the HI has a patchy distribution,  has a distinct signature
where the signal is determined by the size of the discrete ionized 
regions. The signal  at other epochs,  where the HI follows the dark
matter, is determined largely by the power spectrum of dark matter
fluctuations. The signal is strongest for baselines where the antenna
separations are within a few hundred times the wavelength of
observation, and an optimal strategy would preferentially sample these
baselines. In the frequency domain,  for most baselines the
visibilities at two different frequencies 
are uncorrelated beyond $\Delta \nu \sim 1 \, {\rm MHz}$, a signature
which in principle  
would allow the HI signal to be easily distinguished from the continuum
sources of contamination.     

\end{abstract}

\section{Introduction}
One of the major problems in modern cosmology is to determine the
distribution of matter on large scales in the universe and understand
the large scale structure (LSS) formation . The last decade has
witnessed 
phenomenal progress, particularly on the observational front where
CMBR anisotropies (eg. \citealt{spergel}) and galaxy redshift surveys  
(eg. \citealt{tegmarka}, 2003b) in   conjunction with 
other observations have determined the parameters of the background
cosmological model and the power spectrum of density fluctuations to a
high level of precision. These observations are all found to be
consistent with a $\Lambda$CDM cosmological model with a scale
invariant primordial power spectrum, though there are hints of what
could prove to be very interesting features beyond the simplest
model. 

Another very interesting problem, on smaller scales in cosmology, is
the formation of the first gravitationally bound objects and their 
subsequent merger to produce  galaxies.  Observations of quasar
spectra show the diffuse gas in the universe to be completely ionized
at redshifts $z \leq 5$  \citep{fan}. Understanding the reionization
process and its relation with the first generation of luminous objects 
produced by gravitational collapse is one of the challenges facing
modern cosmology.  

Observations of the redshifted 21 cm HI  radiation provide an unique
opportunity for probing the universe from the dark ages when the hydrogen
gas first decoupled from the CMBR to the present epoch. Such
observations will  allow us to probe in detail how the universe was
reionized. In addition, redshifted HI observations will allow the
large scale distribution of matter in the universe to be studies over
a very wide range of redshifts.  The possibility of observing
21 cm emission from the cosmological structure formation  was first
recognized by \citet{suny}. Later studies by \citet{hogan} and
\citet{scott} consider both emission and absorption against the
CMBR.  

\begin{figure}
\includegraphics[width=84mm]{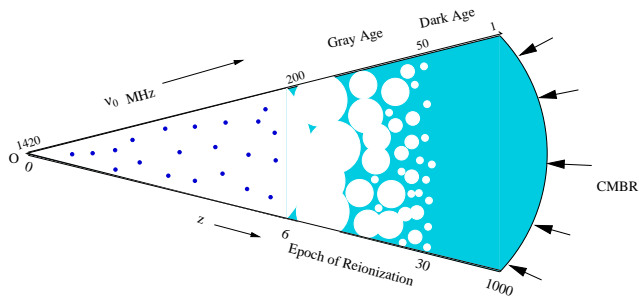}
\caption{This figure schematically shows the evolution of HI (shaded
  in the figure) through   the entire history of the
  universe. Hydrogen decouples from the CMBR   at $z \sim 1000$ when
  electrons and protons combine, for the first   time, to produce
  HI. The subsequent era, until the formation of the    first luminous
  objects at $z \sim 30$, is referred to as the dark 
  ages or the Pre-reionization era. The HI, in this era,  is nearly
  uniformly   distributed with small fluctuations which trace  
  the dark   matter.    The first luminous objects
  reionize the HI. The reionization is initially confined to small
  bubbles surrounding the luminous objects. The bubbles of
  ionized HII gas grow until they finally overlap and the whole
  universe is reionized by $z \sim 6$. The HI distribution in the
  Reionization era or the Gray Age is patchy.  Only the HI in dense
  collapsed objects survive the reionization. The surviving HI clouds
which are seen as DLAs in quasar absorption spectra are shown as
  points in the figure. The large scale distribution of these clouds
  is assumed to follow the dark matter. Also shown is the CMBR which
  propagates through the intervening HI to the observer at $z=0$.} 
\label{fig:a1}
\end{figure}

The cosmological evolution of the HI gas is shown schematically in
Figure 1. The 21 cm emission has been perceived as a very important
tool for studying the epoch of reionization, and the expected 
signature,  in 21 cm,  of the  heating of the HI gas and its
subsequent reionization has been studied extensively
(\citealt{madau};  \citealt{gnedin}; \citealt{shaver};
\citealt{tozzi};  \citealt{isfm}; \citealt{iliev};  \citealt{ciardi}; 
\citealt{furlanetto};  \citealt{gnedin2};  \citealt{miralda};
\citealt{chen}). The timing, duration and character of the events
which led to the reionization of the universe contains an enormous 
amount of information about the first cosmic  structures and
also has important implication for the later generations of baryonic
objects. This process is not very well understood
(eg. \citealt{barkana}),  
but there are  several relevant observational constraints. 

The observation of quasars at redshift $ z \sim 6$ which
show strong HI absorption  \citep{becker} indicates that at least  
$1\%$  of the total hydrogen mass   at $ z \sim 6 $  is neutral
\citep{fan}, and the  neutral mass fraction decreases rapidly   at
lower redshifts. This is a strong indication that the 
epoch of reionization ended at $ z\sim 6 $ .

Observations of the CMBR polarization, generated through Thomson
 scattering of  CMBR photons by free electrons along the line of
 sight, indicates that the reionization began at a redshift $z > 14$.
 On the other hand, the observed anisotropies of the CMBR indicate
 that the total optical  depth of the Thomson scattering is not
 extremely high, suggesting that reionization could not have started at
 redshift much higher than about  30 (\citep{spergel}. 

A third constraint comes from determinations of the IGM temperature
from observations of the ${\rm Ly} \alpha$  forest in the $z$ range
$2$ to $4$ which indicates a complex reionization history with there
possibly being an order unity change in the neutral hydrogen fraction
at $z \le 10$ (\citealt{theuns}; \citealt{hui}). 

The possibility of observing the HI emission at $z \leq 6$, the 
post-reionization era, has also been  discussed extensively
particularly in the context of the Giant Meterwave Radio Telescope
(GMRT) (\citealt{subramanian}; \citealt{kumar}; \citealt{bagla1};
\citealt{bharad1}; \citealt{bagla2}).  These observations will probe
the large scale structure formation at the redshift where the HI
emission originated.

\citet{lz} have recently proposed that observations of the angular
fluctuations in the HI absorptions against the CMBR,  originating in
the $z$ range $30$ to $200$ corresponding to the dark age of the
universe, would allow the primordial power spectrum to be probed at an
unprecedented level of accuracy. \citet{bharad5} have investigated
this in detail  including the effect of redshift space distortion 
and gas temperature fluctuations.

In this paper we discuss the nature and the magnitude of the HI signal
expected across  the entire redshift range starting from the dark age
of the universe at $z \sim 1000$ through the epoch of reionization to
the present era.  Our discussion is in the context of observations
using radio interferometric arrays to study the redshifted HI
signal. Such observations are sensitive to only the angular
fluctuations in the HI signal. The quantity measured in radio
interferometric observations is the complex visibility. Typically, the
visibilities are used to construct an image which is then analysed. An
image of the HI distribution, while interesting, is  not the
optimal tool for quantifying and analysing the redshifted HI
radiation. The quantities of interest are the  statistical properties
of the fluctuations in the redshifted HI radiation, and an image, if
at all useful, is only an intermediate step in the process of
extracting these statistical quantities.   Here we propose
that correlations between the complex visibilities measured at
different baselines and frequencies be used to directly quantify the
angular and frequency domain clustering pattern in the HI radiation,
completely doing away with the need for making an image.    

In Section 2 of this paper we develop the formalism for calculating
visibility correlations, bringing out the relation between this
quantity and various properties of the HI distribution. The
fluctuations, with angle and frequency,  of  the redshifted HI
radiation  can be directly related to fluctuations in the properties
of the HI distribution in space. The mapping is a little complicated
due to the presence of peculiar velocities which moves around spectral
features along the frequency direction. This effect, not included in
most  earlier calculations, is very similar to the redshift space
distortion in galaxy redshift surveys \citep{kais} and has been
included here (also \citealt{bharad1}; \citealt{bharad5}). 

The possibility of using visibility correlations to study the HI
distribution has been discussed earlier in the restricted context of
HI emission from $z <6$ by \citet{bharad2}, \citet{bharad3} and
\citet{bharad4} who have developed the formalism and used it to make
detailed predictions and simulations of  the signal expected at the
GMRT.

\citet{morales} (MH hereafter) have discussed the use of visibility
correlations in  
the context of detecting the HI signal from the epoch  of reionization
using LOFAR.  They also address the issue of extracting the signal
from the  foreground contamination. 

  \citet{zald} (ZFL hereafter) have proposed the use of the angular
  power  
spectrum to quantify the fluctuations in the redshifted HI 
radiation, in direct analogy with the analysis of the CMBR
anisotropies,   As noted by them,  though this analogy is 
partially true in     that  both the redshifted HI radiation and the
CMBR are  background  radiations present at all directions and
frequencies. there is a major difference in that observations at
different frequencies probe the HI  at different distances whereas we
expect exactly the same CMBR anisotropies at all frequencies.  The
  multi-frequency angular power-spectrum introduced by ZFL allows the
cross-correlations in the   angular fluctuations in the HI signal at
  different   frequencies to be quantified. 

The cross correlation between the HI signal at different frequencies
is expected to decay rapidly as the  frequency separation is
increased,  and  as noted by several  earlier authors 
(eg. \citealt{shaver}; \citealt{dimat1}; \citealt{gnedin2};
\citealt{dimat}) this holds the possibility of  
allowing us to distinguish it from the  contaminations which are
expected to have a continuum spectrum and be correlated across
frequencies. ZFL  discuss  this issue focusing on the
properties of the contaminants which could possibly limit the ability
to extract the HI signal, and show that it should be possible to
extract the signal provided  point sources with flux $> 0.1 {\rm mJy}$
can be identified and removed. 

In this paper we build up on the earlier work in this field. In 
particular, we generalize the formalism developed in \citet{bharad2}  
so that it can be used for both,  HI emission and absorption against the
CMBR, thereby extending its scope to the entire redshift range
starting from the epoch when hydrogen  first recombines to the
present. The 
formalism reported here presents progress  over the earlier 
work (MH and ZFL) on at least two counts.
First,  it incorporates the effects of redshift-space distortions, 
  ignored in all earlier works except those by Bharadwaj and
  collaborators (mentioned earlier). As noted by \citet{bharad5},
this is a very important effect and  not  including it changes the signal
by $50 \%$ or more.  Second, our formalism  shows, explicitly,  how the
the correlations in the HI signal decay with increasing frequency
separation $\Delta \nu$.  The $\Delta \nu$ dependence, though implicit
in the formalism of MH and ZFL, has not been explicitly quantified by
these authors. A third point, which in our opinion is also a major
advantage of our formalism over the angular power spectrum (ZFL),  is
that it deals directly with  visibilities which are the primary
quantity  measured in radio-interferometric observations.  Further, 
the visibility correlations have a very simple relation to the
three-dimensional power spectrum  of HI fluctuations, the quantity of
prime interest in HI observations. It may be noted that though the
visibility-correlations and angular power-spectrum are equivalent
descriptors of the redshifted HI observations, the former is simpler
to calculate and interpret,  it being related to the HI  power
spectrum through an exponential function instead of spherical Bessel
functions. 

We next present a brief outline of the paper.  In Section 3, we
trace  the evolution of the cosmological HI from the dark ages to the
present era,  focusing on epochs which are of interest for 21 cm
observations. In Section 4 we present the results for the expected HI
signal from the different epochs,  and  in Section 5 we summarize and
discuss 
our results. 

Finally, we bring to the notice of  readers two recent papers
\citep{morales1} and \citep{santos}, submitted  while this paper
was being revised,  addressing respectively  the power spectrum
sensitivity and the 
$\Delta \nu$ dependence   of of the  epoch of reionization
HI signal.

\section{The formalism for HI visibility correlations}
The story of neutral hydrogen (HI) starts at a redshift $z \sim 1000$ when,
for the first time, the primeval plasma has cooled sufficiently for
protons and electrons to combine. The subsequent evolution of the HI
is very interesting, we shall come back to this once we have the
formulas required to calculate the  HI emission and absorption
against the CMBR. The propagation of the CMBR, from the last
scattering surface where it decouples from the primeval plasma to the
observer at present,  through the intervening neutral or partially
ionized hydrogen is shown schematically in Figure 1. Along any line of
sight,  the  CMBR interacts with the HI  through the spin flip
hyperfine transition at  $\nu_e= c/ \lambda_e=1420 \,{\rm MHz}$
($\lambda_e= 21 \, {\rm cm}$) 
in the rest frame of the hydrogen. This  changes the  brightness 
temperature of the CMBR to  
\begin{equation}
T(\tau)=\Tc e^{-\tau}+T_s(1-e^{-\tau})
\label{eq:a1}
\end{equation}
where $\tau$ is the optical depth,  $\Tc$ is the background
CMBR temperature  and $T_s$ is the spin temperature defined
through the level population ratio 
\begin{equation}
\frac{n_1}{n_0}=\frac{g_1}{g_0}  \exp(-T_*/T_s) \,. 
\label{eq:a2}
\end{equation}
Here $n_0$ and $n_1$ refer respectively to  the number density of
hydrogen atoms in the ground  and excited states of the hyperfine
transition, $g_0=1$ and $g_1=3$ are  
the degeneracies of these levels and $T_*=0.068 \, K = h_p \, \nu_e/
k_B$,   where $h_P$ is the Planck constant and $k_B$ the Boltzmann
constant.

The absorption/emission features introduced by the intervening HI are
redshifted to a frequency $\nu=\nu_e/(1+z)$ for an observer at
present.  
The expansion of the universe, and  the HI  peculiar velocity 
both   contribute to the redshift.  Incorporating  these effects,  the
optical depth at a redshift $z$ along a line of sight $\n$ is
(Appendix A)  
\begin{eqnarray}
\tau &=& \frac{4.0 \, \mK }{\Ts}\, \left(\frac{\Omega_b
  h^2}{0.02}\right) 
\left( \frac{0.7}{h} \right) \frac{H_0}{H(z)} (1+z)^3 \,  \nonumber \\ 
&\times&   \frac{\rho_{\HI}}{\bar{\rho_{\H}}} \,
  \left[1-\frac{(1+z)}{  
  H(z)}\frac{\partial v}{\partial \rn}\right]    
\label{eq:a3}
\end{eqnarray}
where  $r_{\nu}$ and $H(z)$  are  respectively the comoving distance
and the Hubble  parameter, as functions of $z$ or equivalently  $\nu$,
calculated  using the background cosmological model with no peculiar
velocities. Also,  $\rho_{\HI}/\bar{\rho}_{\H}$ is the ratio 
of the neutral  hydrogen to the mean hydrogen density,  and   $v$ is
the line of sight component of the peculiar velocity of the HI, both
quantities being evaluated at a comoving position  $\x=\rn  \n$.  The
optical depth (eq. \ref{eq:a3}) is  typically much less than unity.  

The quantity of interest for redshifted $21 \, {\rm cm}$
observations,  the excess brightness temperature  redshifted to the
observer at present is 
\begin{equation}
\delta T_b(\n,z) =\frac{T(\tau)-\Tc}{1+z} \, \approx
\frac{(T_s-\Tc)\tau}{1+z}  
\label{eq:a4}
\end{equation}

It is convenient to express the  excess brightness temperature as 
$\delta \Tb(\n,z) = \bar{T}(z) \times \d(\n,z)$
where 
\begin{equation}
\bar{T}(z)=4.0 \, \mK   (1+z)^2  \, \left(\frac{\Omega_b
  h^2}{0.02}\right)  \left(\frac{0.7}{h} \right) \frac{H_0}{H(z)}   
\label{eq:a5}
\end{equation}
depends only on $z$ and  the background cosmological parameters,
and 
\begin{equation}
\d(\n,z)= \frac{\rho_{\HI}}{\bar{\rho}_{\H}} \left(1-\frac{\Tc}{\Ts}
 \right)  \left[1-\frac{(1+z)}{  H(z)}\frac{\partial v}{\partial
 r}\right]    
\label{eq:a6}
\end{equation} 
is the ``$21 {\rm cm}$ radiation  efficiency''
$(1-\Tc/\Ts)\, \rho_{\HI}/\bar{\rho}_{\H}$ introduced by
\citet{madau}, with  the extra velocity  term arising  here  on
account of  the  HI  peculiar velocities. We refer to $\d$ as the ``21
cm radiation 
efficiency in redshift space''. This  varies 
with position $(\x= \rn \n)$ and redshift, and it  incorporates the
details of 
the HI evolution and the effects of the growth of large scale
structures. We also
introduce $\dt(\k,z)$, the Fourier transform of  $\d(\x,z)$, 
defined through  
\begin{equation}
\d(\n,z)=\int \frac{d^3 k}{(2 \pi)^3} e^{-i \,  \k \cdot \rn \n} 
 \dt(\k,z)  \, .
\label{eq:a9}
\end{equation}
and the associated three dimensional power spectrum $P_{\HI}(\k,z)$  defined through 
\begin{equation}
\langle \dt(\k,z) \, \dt^{*}(\k^{'},z) \rangle = (2 \pi)^3 \,  
\delta^3_D(\k-\k^{'}) \, P_{\HI}(\k,z)
\label{eq:a13}
\end{equation}
 where $\delta^3_D$ is the three dimensional Dirac delta function.

When discussing  radio interferometric observations, it is  
convenient to use the specific intensity $I_{\nu}$ instead of the
brightness temperature. In the Raleigh-Jeans limit $\delta
I_{\nu}  =[2 \, k_B/\lambda_e^2 (1+z)^2] \, \times \, \delta T_b
=\bar{I}_{\nu} \times \d$ 
where 
\begin{equation}
\bar{I}_{\nu}=2.5 \times 10^2 \, \frac{\rm Jy}{\rm sr}  \,
  \left(\frac{\Omega_b   h^2}{0.02}\right)  \left(\frac{0.7}{h}
  \right) \frac{H_0}{H(z)}    
\label{eq:a7}
\end{equation}
The quantities  $\bar{T}$ and $\bar{I}_{\nu}$ are determined by 
cosmological parameters whose values are reasonably well
established. In the rest of the paper, when required,  we shall use
equations  (\ref{eq:a5}) and (\ref{eq:a7}) with
$(\Omega_{m0},\Omega_{\lambda0},  \Omega_b h^2,h)=(0.3,0.7,0.02,0.7)$
to calculate $\bar{T}$ and $\bar{I}_{\nu}$.  The crux of the issue of  
observing  HI is in the evolution $\d$ which  is largely
unknown.  We shall discuss a possible scenario in the next section,
before that we discuss  the role of $\d$ in radio interferometric
observations. 

We consider radio interferometric observations using an array of low
frequency radio antennas distributed  on a plane. The  antennas all
point in the same direction ${\bf m}$ which 
we take to be  vertically up wards.  The beam pattern  
$A(\theta)$ quantifies how the  individual antenna, pointing up wards,
responds to signals from different directions in the sky. This is
assumed to be a Gaussian $A(\theta)=e^{-\theta^2/\theta_0^2}$ with 
$\theta_0 \ll 1$ {\it i.e.} the beam width of  the antennas is small,
and the part of the sky which contributes to the signal can be  well
approximated by a plane.  In this approximation the unit vector $\n$
can be represented by $\n={\bf m}+\th$, where $\th$ is a two
dimensional vector in the plane of the sky. Using this $\delta
I_{\nu}$  can be expressed as  
\begin{equation}
\delta I_{\nu}(\n)= \bar{I}_{\nu} \int \frac{d^3 k}{(2 \pi)^3} e^{-i
 \,\rn \, (\kp+  \kpr  \cdot \th)}  
 \dt(\k,z)  \, 
\label{eq:a12}
\end{equation} 
where $\kp=\k \cdot {\bf m}$ and $\kpr$ are respectively the
components of $\k$ parallel and perpendicular to  ${\bf m}$. The
component $\kpr$ lies in the plane of the sky. 

The quantity  measured in interferometric  observations is the complex  
visibility  $V(\u,\nu)$ which is recorded for every independent pair
of antennas at every frequency channel in the band of
observations. For any pair of antennas, $\u={\bf d}/\lambda$
quantifies the separation ${\bf d}$ in units of the wavelength
$\lambda$, we  refer to this dimensionless quantity $\u$ as a
baseline. A typical radio interferometric array simultaneously 
measures visibilities at a large number of baselines and frequency
channels. Ideally, each visibility records a single mode of the
Fourier transform of the specific intensity distribution
$I_{\nu}(\th)$ on the sky.  In reality, it is the Fourier transform of
$A(\theta) \, I_{\nu}(\th)$ which is recorded 

\begin{equation}
V(\u,\nu)= \int d^2 \theta  A(\th) \,  I_{\nu}(\th) \, e^{- i
2 \pi \u \cdot \th} \, .
\label{eq:a8}
\end{equation}

The measured visibilities are Fourier inverted to determine  the
specific intensity distribution $I_{\nu}(\th)$,  which is what we 
call an image. Usually the visibility at zero spacing  $\u=0$ is not
used, and an uniform specific intensity distribution make no
contribution to the visibilities.  The visibilities record only the
angular fluctuations in $I_{\nu}(\theta)$. If the specific intensity
distribution on the sky were decomposed into Fourier modes, ideally
the visibility at a baseline $\u$ would record the complex amplitude
of only a single mode whose period is  $1/U$ in angle on the sky and
is oriented in the  direction parallel to $\u$. In reality, the
response to Fourier modes on the sky is smeared a little because of
$A(\theta)$, the antenna beam pattern,  which appears in
equation (\ref{eq:a8}). 

We now calculate the visibilities arising from  angular fluctuations
in the 
specific intensity excess (or decrement) $\delta I_{\nu}(\th)$
produced by redshifted HI emission (or absorption) against
the CMBR. Using eq. (\ref{eq:a12}) in eq. (\ref{eq:a8}) gives us 
\begin{equation}
V(\u,\nu)=\bar{I}_{\nu} \int \frac{d^3 k}{(2 \pi)^3} a(\u -
  \frac{\rn}{2 \pi} \kpr ) \dt(\k,z) e^{-i \kp \rn}
\label{eq:a11}
\end{equation} 

where  $a(U)$ the Fourier transform of the antenna beam pattern
$A(\theta)$ 
\begin{equation}
A(\theta)=\int d^2 U \, e^{-i \, 2 \pi \u \cdot \th} \,  a(U) \,. 
\label{eq:a10}
\end{equation}
For a Gaussian beam $A(\theta)=e^{-\theta^2/\theta^2_0}$,  the Fourier
transform also is a  Gaussian 
$a(\u)=\pi \theta_0^2 \exp{\left[-\pi^2 \theta^2_0 U^2 \right]}$ 
which we use in the rest of this paper. 

In equation  (\ref{eq:a11}), 
the contribution to  $V(\u,\nu)$ from different modes  is peaked
around the values of $\k$ for which  
  $\kpr  = 2 \pi \u/\rn$. This is so because  the  visibility
  $V(\u,\nu)$ responds to angular fluctuations of period $1/U$ on the
  sky, which corresponds to a spatial period $\rn/U$ at the  distance 
 where the HI is located. It then follows that $\dt(\k,z)$ will
  contribute to a visibility $V(\u,\nu)$ only if the component of $\k$
  projected on the plane of the sky satisfies $\kpr = 2 \pi
  \u/\rn$.  The effect of the
  antenna beam  pattern  is to introduce a range in $\kpr$
  of   width $\mid \Delta  \kpr \mid \sim 2 / \rn \theta_0$ to
  which   each visibility responds.  

We use eq. (\ref{eq:a11}) to calculate $\langle V(\u,\nu)
V^{*}(\u^{'},\nu+\Delta \nu)\rangle$,  the  correlation expected 
between the  visibilities measured  at two different baselines $\u$
and $\u^{'}$, at two frequencies $\nu$ and $\nu+\Delta \nu$ which are 
slightly different {\it ie.} $\Delta \nu/\nu \ll 1$. The
change $\nu \rightarrow \nu + \Delta \nu$ will cause a  very small
change in all the terms  in eq. (\ref{eq:a11}) except for the phase
which we can 
write as $ e^{-i \, \kp (\rn  + \rnp \Delta \nu)} $ where $\rnp=d
\rn/d \nu$. Using this the visibility correlation is 
\begin{eqnarray}
\langle V(\u,\nu) && V^{*}(\u^{'},\nu+\Delta \nu)\rangle = \bar{I}^2_{\nu}
\int \frac{d^3 k}{(2 \pi)^3} \, a( \u - \frac{\rn}{2 \pi} \kpr 
) \nonumber \\ && a^{*}( \u^{'} - \frac{\rn}{2 \pi} \kpr )
P_{\HI}(\k) e^{i \kp \rnp \Delta \nu}
\label{eq:a14}
\end{eqnarray}
The visibilities at $\u$ and $\u^{'}$ will be correlated only if there
 is a significant overlap between the terms  
 $a( \u - \frac{\rn}{2 \pi} \kpr )$ and $a^{*}(\u^{'} -
 \frac{\rn}{2 \pi} \kpr)$ which are peaked around different values
 of $\kpr$. It then follows that  visibilities at two  different
 baselines are correlated only if $\mid \u-\u^{'}\mid < 1/(\pi
 \theta_0)$ and  the visibilities at widely separated baselines are 
 uncorrelated.  To understand the nature of the visibility
 correlations it will suffice  to restrict our analysis to  
$\u=\u^{'}$, bearing in mind that the correlation falls off very
quickly if $\mid \u-\u^{'} \mid  > 1/(\pi \theta_0)$.  A further
simplification is possible if we assume that $U \gg 1/ 2 \pi\theta_0$
 for which we can approximate the Gaussian $a(U)$ with a Dirac delta
 function using $a^2(\mid \u-(\rn /2 \pi) \kpr \mid) \approx (2  
\pi^3 \theta_0^2/\rn^2) \delta^2_D(\kpr - (2 \pi /\rn) \u)$ whereby 
the integrals over $\kpr$ in equation (\ref{eq:a14}) can be
evaluated. We then have  
\begin{equation}
\langle V(\u,\nu)  V^{*}(\u,\nu+\Delta \nu)\rangle =
\frac{\bar{I}^2_{\nu} \theta_0^2}{4 \rn^2}
\int_{-\infty}^{\infty} d \kp \, P_{\HI}(\k) \, e^{i \kp \rnp \Delta \nu}
\label{eq:a16}
\end{equation} 
where $\k=\kp {\bf m} + (2 \pi/\rn) \u$. We expect $P_{\HI}(\k)$ to be
 an even function of $\kp$. It then follows that the imaginary part of
 the visibility correlation  is zero , and the real part is 

\begin{eqnarray}
\langle V(\u,\nu) V^{*}(\u,\nu+\Delta \nu)\rangle &=& 
\frac{\bar{I}^2_{\nu} \theta_0^2}{2 \rn^2}  \label{eq:a17} 
 \\ &\times  & \int_0^{\infty}
d \kp \, P_{\HI}(\k)   \nonumber 
\cos(\kp \rnp \Delta \nu)
\end{eqnarray}

Considering the case $\Delta \nu=0$ first, we see that  the visibility 
correlation $\langle V(\u,\nu) V^{*}(\u,\nu) \rangle$ is proportional
to $P_{\HI}(\k)$ integrated over $\kp$. This directly 
reflects the fact that a baseline $\u$ responds to all Fourier modes 
$\k$ whose projection in the plane of the sky matches $\kpr=(2
\pi/\rn) \u$. A mode $\k=\kp {\rm m} + (2\pi/\rn) \u$ satisfies this
condition for arbitrary values of $\kp$ which is why we have to sum up
the contribution from all values of $\kp$ through the
integral. Further interpretation is simplified if we assume that
$P_{\HI}(\k)$ is an isotropic, power law  of $k=\sqrt{\kp^2+(2 \pi  
  U/\rn)^2}$ {\it ie.} $P_{\HI}(k) \propto k^{n}$. The integral
converges for $n<-1$ for which we have  
$\langle V(\u,\nu) V^{*}(\u,\nu) \rangle \propto  U^{1+n}$  and it is
a  monotonically decreasing function of $U$, {\it ie.} the correlations
will always be highest at the small baselines. This  behaviour
holds in much of the range of $\rn$ and $U$ which we are interested
in. In general $P_{\HI}(\k)$ will not be an isotropic
function of $\k$, but the qualitative features discussed above will
still hold. 

The correlation between the visibilities falls as $\Delta \nu$ is
increased. To estimate $\Delta \nu$ where the visibility correlation
falls to zero,  we assume that  $P_{\HI}(\k) \propto k^{n}$ with
$n<0$. The integral in eq. (\ref{eq:a17})  will have a substantial
value if the power spectrum  falls significantly before the 
term $\cos(\kp \rnp \Delta \nu)$  can complete one oscillation, and
the integral will vanish if the power spectrum does not fall fast
enough. Using the fact that the power spectrum will change by order
unity when $\Delta P_{\HI}/P_{\HI} \sim n \Delta \kp/(2 \pi U/\rn) \sim
1$ and that the $\cos$ term completes  a full oscillation when
$\Delta \kp \rnp \Delta \nu \sim 2 \pi$, we have an estimate $\Delta
\nu \sim \rn/(\rnp U)$ for the value of $\Delta \nu$ beyond which the
visibilities become uncorrelated.  The point to note is that the fall
in the visibility correlation $\langle V(\u,\nu) V^{*}(\u,\nu+\Delta
\nu) \rangle$ with increasing $\Delta \nu$ depends on the combination 
$\rn/\rnp$ which in turn depends only on $z$ and the cosmological
parameters. This holds the possibility of using HI  observations to
probe the background cosmological model over a large range of  $z$.   

It may be noted that we have assumed infinite frequency resolution
when developing the formalism. Actual observations will have frequency
channels of finite width. We briefly discuss the implications of this
in Section 5. 

In the next section we shall consider a  model for the evolution of
the HI, and use this to make predictions for the visibility
correlations from redshifted 21 cm radiation. 

\section{The evolution of the HI}
The evolution of the HI is shown schematically in Figure 1. We 
identify  three different eras namely  Pre-reionization, Reionization
and Post-reionization and discuss these separately. 
\subsection{Pre-reionization}
We resume the story of HI at the redshift $z \sim 1000$ where it
decouples from the CMBR and the universe enters the Dark
Age. Subsequent to this the HI gas temperature $\Tg$ is maintained  at
$\Tc$    by the CMBR which pumps energy into the gas  through the
scattering of  CMBR photons by the small fraction of 
surviving free electrons.    This process becomes ineffective in
coupling $\Tg$ to $\Tc$ at $z \sim 200$. In the absence of external
heating  at $z<200$ the gas cools adiabatically with  $\Tg \propto
(1+z)^2$ while   $\Tc \propto (1+z)$. The spin temperature $\Ts$ is
strongly coupled to $\Tg$ through the collisional spin flipping
process until $z \sim 70$. The collisional process is weak at lower
redshifts, and $\Ts$ again approaches $\Tc$.  This gives a window in
redshift $30 \le z \le 200$ where   $\Ts < \Tc$ and $\d <0$. The HI in
this redshift range will produce absorption features in the CMBR
spectrum. \citet{lz} have proposed that observations of the angular
fluctuations in this signal holds the promise of allowing the
primordial power spectrum of density perturbations to be studied at an
unprecedented level of accuracy. The nature of this signal was studied
in detail in \citet{bharad5}, incorporating the effects of peculiar
velocities and gas temperature fluctuations. 

We assume that all the hydrogen is neutral,  and that the fluctuations
in the HI density $\Delta \rho_{\HI}$  trace the dark matter
fluctuations $\Delta$  {\it ie}  $\Delta\rho_{\HI}(\k,z)=
\bar{\rho}_{\H} \Delta(\k,z)$. Th HI peculiar velocities are also
determined by the dark matter,  and the Fourier transform of
$[(1+z)/H(z)] (\partial v/\partial r)$ in equation (\ref{eq:a6}) is
$-f (\Omega_m) \mu^2\Delta(\bf{k},z)$, where  $\mu$ is the cosine of
the angle between  ${\bf k}$ and the line of sight $\n$, and
$f(\Omega_m) \approx \Omega_m^{0.6}+\frac{1}{70} \left[1-\frac{1}{2}
  \Omega_m (1+\Omega_m) \right]$  in a spatially flat universe
\citep{lahav}. The fluctuations
$\Delta\rho_{\HI}$ also cause  fluctuations in the spin temperature
$\Delta \Ts$ which we 
quantify using a dimensionless function  $s(z)$ defined such that
$\Delta \Ts=s\,  \Ts \, \Delta \rho_{\HI}/\rho_{\HI}$. Using these in
equations (\ref{eq:a6}) and (\ref{eq:a9}) we have the Fourier transform
of the ``21 cm radiation efficiency in redshift space'' 
\begin{equation}
\dt(\k,z) =
 \left[ \left(1 - \frac{\Tc}{\Ts}\right) \left(1+ f \mu^2 
  \right)  +  \frac{\Tc}{\Ts} s \right] \Delta({\bf k},z) \, 
\label{eq:b1}
\end{equation}
which we use to calculate $P_{\HI}$, the power spectrum of the
radiation efficiency  in terms of the dark matter power spectrum
$P(k)$ 
\begin{equation}
P_{\HI}(\k,z)= \left[ \left(1 - \frac{\Tc}{\Ts}\right) \left(1+
  f \mu^2  
  \right)  +  \frac{\Tc}{\Ts} s \right]^2 P(k,z) \,.
\label{eq:b2}
\end{equation}
Observations of the angular fluctuations in the HI radiation using
either the visibility correlations or other methods like the angular
power spectrum will determine $P_{\HI}$ (eq. \ref{eq:a17}). These
observations  will directly probe  the  matter power spectrum provided
the evolution of $(1-\Tc/\Ts)$ and $s \, \Tc/\Ts$ are known. This has
been investigated in detail in \citet{bharad5} and we use the results
here. The quantities $(1-\Tc/\Ts)$ and $s \, \Tc/\Ts$ also
determine the frequencies where we expect a substantial   signal. 

We use eq. (\ref{eq:b2}) to calculate the visibility correlations
expected from the Pre-reionization era. 
\subsection{Reionization}
The Dark Age of the universe ended,  possibly at a redshift $z \sim
30$, when the first luminous objects were formed.  The radiation from
these luminous objects and from the subsequently formed luminous
objects   ionized  the low density HI in the universe. 
Initially only small spherical regions (Stromgren sphere) surrounding
the luminous 
objects  are filled with ionized HII gas,  the rest of the universe
being filled with HI (Figure 1).  Gradually 
these spheres of ionized gas grow until they finally  overlap, filling
up the whole of space, and all the low density gas in the universe is 
ionized.  Here we adopt a simple model for the reionization
process, The model, though simple, is sufficiently general to
reasonably describe  a variety of scenarios by changing
the values of the model parameters. 

We assume that the HI gas is heated well before it is reionized, and
that the spin temperature is coupled to the gas temperature with 
$\Ts \gg \Tc$ so that  $(1-\Tc/\Ts) \rightarrow 1$. It then follows
that $\d >0$  (eq. \ref{eq:a6}) {\it ie.} the HI will be seen in
emission. Also, the HI 
emission depends only on the HI density and peculiar velocity.  
Like in the Pre-reionization era, we assume that the hydrogen
density and peculiar velocities follow the dark matter, the only 
difference being that a fraction of the volume $f_V$ is now 
ionized. We assume that non-overlapping spheres of comoving radius $R$
are completely ionized,  the centres of the  spheres being clustered
with a bias $b_c \geq 1$ relative to underlying   dark matter
distribution. This model is  similar to 
that used by \citet{zald} in the context of HI emission,  and
\citet{gruz} and \citet{knox} in the context of the effect of patchy
reionization on the CMBR.  There is a difference in that we have
assumed the centers of the spheres to be clustered, whereas Zaldarriaga
et al. (2003) assumed them to be randomly distributed. One would
expect the centres of the ionized spheres to clustered,  given the fact
that we identify them with the locations of the first luminous objects
which are believed to have formed at the peaks of the density
fluctuations.  

\begin{figure}
\includegraphics[width=84mm]{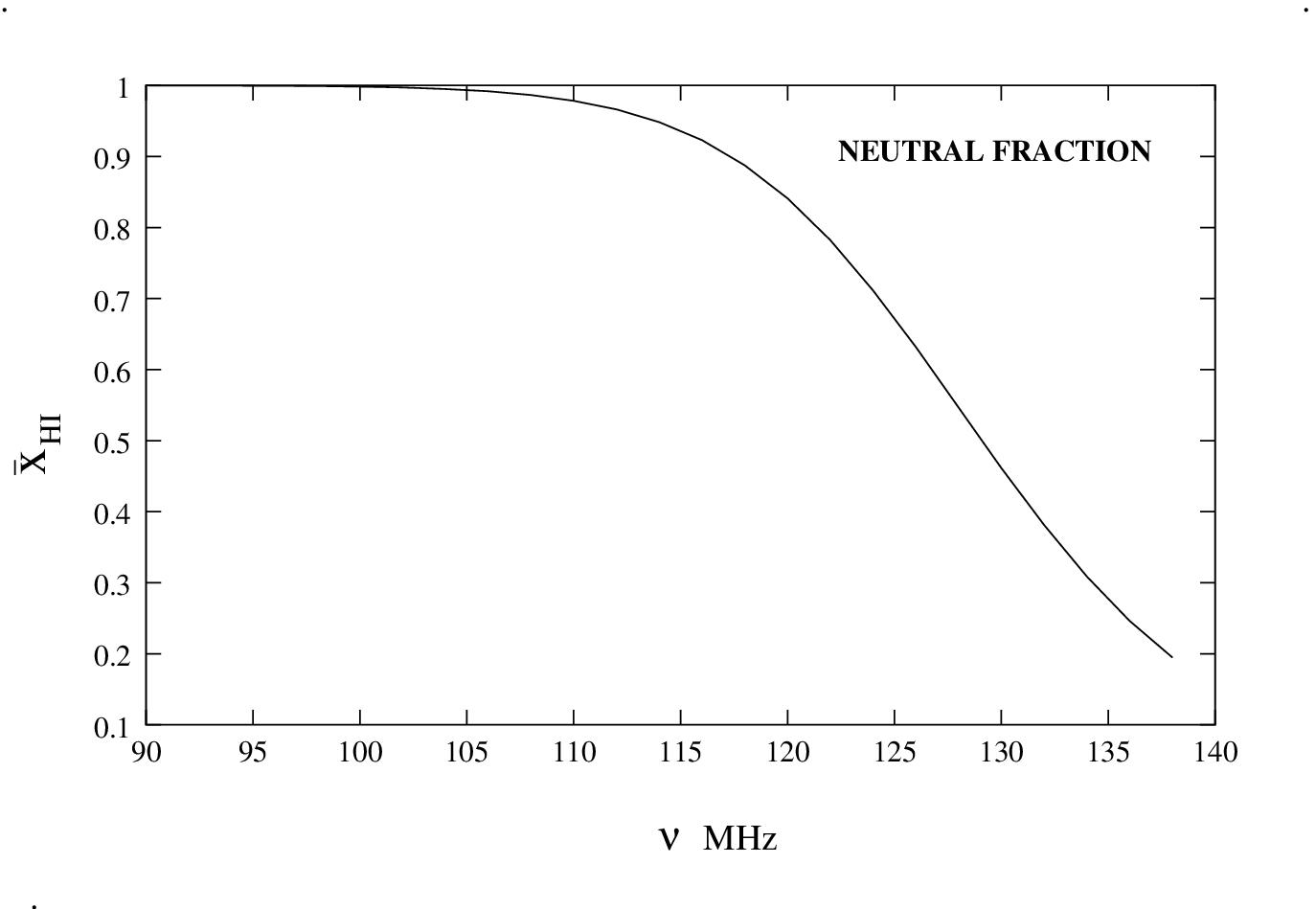}
\caption{This shows the evolution of the mean neutral fraction
  $\bar{x}_{\HI}$ with the frequency of the redshifted HI emission
  for  the reionization model discussed in the text.}
\label{fig:a2}
\end{figure}

 The  fraction of volume ionized $f_V$, the   mean neutral fraction
 $\bar{x}_{\HI}$   and mean comoving number density of ionized 
 spheres $\bar{n}_{\HI}$ are related  as $f_V=1-\bar{x}_{\HI}=(4 
 \pi  R^3/3) \bar{n}_{\HI}$. Following Zaldarriaga et al. (2003),  we 
 assume that   
\begin{equation}
\bar{x}_{\HI}(z)=\frac{1}{1+\exp((z-z_0)/\Delta z)}
\end{equation}
with $z_0=10$ and $\Delta z=0.5$ so that $50 \%$ os the hydrogen is
reionized at a redshift $z=10$ (Figure \ref{fig:a2}). The comoving
size 
of the spheres $R$ increases  with $z$ so that it is $3 \,h^{-1} {\rm
  Mpc}$ at $z=10$, and  $R$ increases in such a way so that the mean
comoving number density of the  ionized spheres  is constant at
$\bar{n}_{\HI}= 4.4 \times 10^{-3} \, h^{3} {\rm   Mpc}^{-3}$.  

In this model the HI density is  $\rho_{\HI}(\x,z)=\bar{\rho}_{\H} (1
+ \delta) \left[ 1 - \sum_a \theta(\mid \x -\x_a \mid/R) \right] $
where $\delta$ is the dark matter fluctuation, $a$ refers to the
different ionized spheres with centers at $\x_a$, and $\theta(y)$ is the
Heaviside step function defined such that $\theta(y)=1$ for $0 \le y
\le 1$ and zero otherwise. We then have 
\begin{equation}
\d(\x,z)=\left[ 1+\delta-\frac{1+z}{H(z)} \frac{\partial v}{\partial r}
  \right]  \left[ 1 - \sum_a \theta(\frac{\mid \x -\x_a \mid}{R})
  \right] \,. 
\label{eq:c1}
\end{equation}
We have already discussed the Fourier transform of the first part of
eq. (\ref{eq:c1}) which is $(2 \pi)^3 \delta^3_D(\k)+(1+\mu^2) 
\Delta(\k,z)$,  and the Fourier    transform of the second part is $(2
\pi)^3 \delta^3_D(\k) - (4 \pi R^3/3) W(k R) \sum_a e^{i \k \cdot
  \x_a}$ where $W(y)=(3/y^3)[\sin(y) - y \cos(y)]$ is the spherical
top hat window function.  The term  $\sum_a e^{i \k 
  \cdot \x_a}$ is the Fourier transform of the distribution of the
centers of the ionized spheres which we  write as $\sum_a e^{i \k
  \cdot \x_a}= \bar{n}_{\HI} [(2 \pi)^3 \delta^3_D(\k) +
  \Delta_P(\k,z) + b_c  \Delta(\k,z) ]$, where $\Delta_P(\k,z)$ is the
fluctuation in the distribution of the centers  arising  due to the
discrete nature of these points (Poisson fluctuations) and  the term
$b_c \Delta(\k,z)$ arises due to the clustering of the centers. These two 
components of the fluctuation are independent and $\langle
\Delta_P(\k) \Delta^{*}_P(\k^{'}) \rangle = (2 \pi)^3 \delta^3_D(\k -
\k^{'}) \times 1/\bar{n}_{\HI}$. Convolving the Fourier transform of
the two terms in eq. (\ref{eq:c1}), and dropping terms of order
$\Delta^2$ and $\Delta \Delta_P$ we have 
\begin{eqnarray}
\dt(\k,z)&=&\left[\bar{x}_{\HI} (1+ f \mu^2) - b_c f_V W(k R) \right]
\Delta(\k,z)  \nonumber \\
&-& f_V W(k R) \Delta_P(\k,z) 
\label{eq:c2}
 \end{eqnarray}
This gives the power spectrum of the ``21 cm emission efficiency in
redshift space'' to be 
\begin{eqnarray}
P_{\HI}(\k,z)&=&\left[\bar{x}_{\HI} (1+ f \mu^2) - b_c f_V W(k
  R) \right]^2  P(k,z)  + \nonumber \\
&+& \frac{f^2_V W^2(k R) }{\bar{n}_{\HI}} 
\label{eq:c3}
 \end{eqnarray}
The first term which contains $P(k)$ arises  from the clustering of
the hydrogen and the clustering of the centers of the ionized
spheres. The second term which has $1/\bar{n}_{\HI}$  arises due to
the discrete nature of the ionized regions. 

Our model has a limitation that it cannot be used  when a large
fraction of the volume is ionized as the ionized spheres start to
overlap and the HI density becomes negative in the overlapping
regions. Calculating  the fraction of the total volume where the HI
density is negative, we find this to be $ f_V^2/2$   under the
assumption that the   centers of the ionized spheres are randomly 
distributed.  We use this to asses the range of validity of our 
model. We restrict the model to $z>10$ where $f_V<0.5$, and the HI
density is negative in less than $12.5 \%$ of the total volume. 
The possibility of the spheres overlapping  increases if
they are highly clustered,  and we restrict $b_c$ to 
$b_c =1.5 $  throughout to keep this under control.  

We use equation (\ref{eq:c3}) to calculate the visibility correlations
during the Reionization era. 

\subsection{Post-reionization}
All the low density hydrogen is ionized by a redshift $z \sim 6$ and
HI is to be found only in high density clouds (Figure 1), possibly
protogalaxies,  which survive  reionization.   Observations of
Lyman-$\alpha$ 
absorption lines  seen in quasar spectra have been used to determine 
the HI density in the redshift range $1 \le z \le 3.5$. 
These observations currently indicate $\Omega_{gas}(z)$, the comoving
density of neutral gas expressed as a fraction of the present critical 
density, to be  nearly constant at a value   $\Omega_{gas}(z) \sim
10^{-3}$ for   $z \ge 1$ \citep{peroux}. 
The bulk of the neutral gas is in clouds which have  HI column   
densities greater than $2 \times 10^{20} {\rm atoms/cm^{2}}$
(\citealt{peroux}, \citealt{lombardi}, \citealt{lanzetta}). These
high column density clouds are responsible for 
the damped Lyman-$\alpha$ absorption lines  observed  along lines  of
sight to quasars. The flux of HI emission from individual clouds
($ < 10 \mu {\rm Jy}$) is too weak to be detected by existing
radio telescopes unless the image of the cloud is   significantly
magnified  by an intervening cluster gravitational lens
\citep{saini}. 

For the purposes of this paper, we assume that the HI clouds trace the
dark matter.  It may be noted that  we do not really
expect this to be true,  and  the HI clouds clouds will, in all
probability, be biased with respect to the dark matter. However, our
assumption is  justified given the fact that we currently have very
little information about the actual distribution of the HI  clouds, 

Converting $\Omega_{gas}$ to the mean neutral fraction
$\bar{x}_{\HI}=\bar{\rho}_{\HI}/\bar{\rho}{\H}=\Omega_{gas}/\Omega_b$
gives us $ \bar{x}_{\HI}=50 \Omega_{gas}  h^2 (0.02/\Omega_b h^2)$ or
$\bar{x}_{\HI}=2.45 \times 10^{-2}$. We also assume $\Ts \gg \Tc$ and
hence we see the HI in emission.   Using these we have 
\begin{equation}
P_{\HI}(\k,z)=\bar{x}^2_{\HI}\left( 1+ f  \mu^2 \right)^2
P(k,z) 
\label{eq:d1}
\end{equation}
The fact that the neutral hydrogen is in discrete clouds  makes a
contribution which we do not include here.  This effect originates
from the fact that the HI emission line from individual clouds  has a
finite width, and the visibility correlation is enhanced when $\Delta
\nu$ is smaller than the line-width of the emission from the individual
clouds. Another important effect not included here is that the
fluctuations become non-linear at low $z$. Both these effects have
been studied using simulations \citep{bharad4}, The simple analytic
treatment adopted here suffices for the purposes of this paper where
the main focus is to estimate the magnitude and the nature of the HI
signal over a very large range of redshifts starting from the Dark
Ages to the present.

\section{Results }

In this section we present predictions for the visibility correlation
signal from the different eras.  Equation (\ref{eq:a17}) allows us to
calculate the visibility correlation by evaluating just a single
integral. This is valid only if $U \gg (2 \pi \theta_0)^{-1}$,  failing
which we  have to  use eq. (\ref{eq:a14}) which involves a three
dimensional integral. \citet{bharad3} have compared the results
calculated using  equations  (\ref{eq:a14})  and (\ref{eq:a17}),  and
find that they agree quite well over the range of $U$ of interest. 
Here we use eq. (\ref{eq:a17}) throughout.  

In equation (\ref{eq:a17}),  it is necessary to specify $\theta_0$
which is the size of the beam of the individual antennas in
the array.  It may be noted that $\theta_0 \approx 0.6 \times
\theta_{\rm FWHM}$. The value of $\theta_0$ will depend on the
physical dimensions of the antennas and the wavelength at which the
observations are being carried out. We have assumed that
$\theta_0=1^{\circ}$ at $325 {\rm MHz}$ and scaled this  as
$\theta_0 \propto \lambda$ to determine the value at other
frequencies. This value for $\theta_0$ is appropriate for the GMRT
which can operate at frequencies down to $150 {\rm MHz}$. 
The visibility correlations scale as $\theta_0^2$, and it
is straightforward to scale the results presented here to make
 visibility correlation predictions for  other radio telescopes. 

The other input needed to calculate visibility correlations is
$P_{\HI}(\k)$,  the power spectrum of the ``21 cm radiation efficiency
in redshift space''. In the models for the three eras discussed in the
previous section, the only quantity which is left to be specified is
the dark matter power spectrum $P(k)$ which we take to be that of the
currently favoured  $\Lambda$CDM  model with the shape parameter
$\Gamma=0.2$, normalized to $\sigma_8=1$ at present. We use equations
(\ref{eq:b2}),  (\ref{eq:c3}) and (\ref{eq:d1}) for $P_{\HI}(\k)$ in
the pre-reionization, reionization and post-reionization eras
respectively. 

\begin{figure}
\includegraphics[width=84mm]{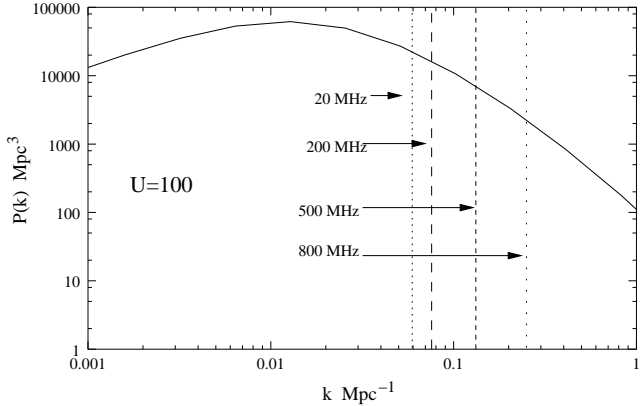}
\caption{This shows the linear power spectrum of dark matter   density  
  fluctuations used to calculate the expected HI visibility
  correlations. The power spectrum is shown at the present epoch. 
A baseline $\u$ at a frequency $\nu$ will probe the power spectrum at
  all Fourier modes $k \ge k_{min}=(2 \pi / \rn) U$. The values of
  $k_{min}$ are shown at different frequencies  for $U=100$. This
  figure can be used to determine the length-scale probed by any 
  baselines $U$. 
}\label{fig:a3}
\end{figure}

The visibility correlation $\langle V(\u,\nu) V^{*}(\u,\nu+\Delta \nu)
\rangle$ responds to the power spectrum at Fourier modes $k \ge
k_{min}=(2  \pi/\rn)U$.  Figure \ref{fig:a3}  shows the  part of the
power 
spectrum which will be probed at different frequencies  by the
visibility correlations at the baseline $U=100$. The point to note is
that a typical radio interferometric array with a large range of
baselines and frequencies will provide many  independent measurements
of the power spectrum at different epochs and in overlapping bins of
Fourier modes.  This will allow the evolution of the HI and the dark
matter power  spectrum to be simultaneously studied at a high level of
precision.

\subsection{Pre-reionization}
\begin{figure}
\includegraphics[width=84mm]{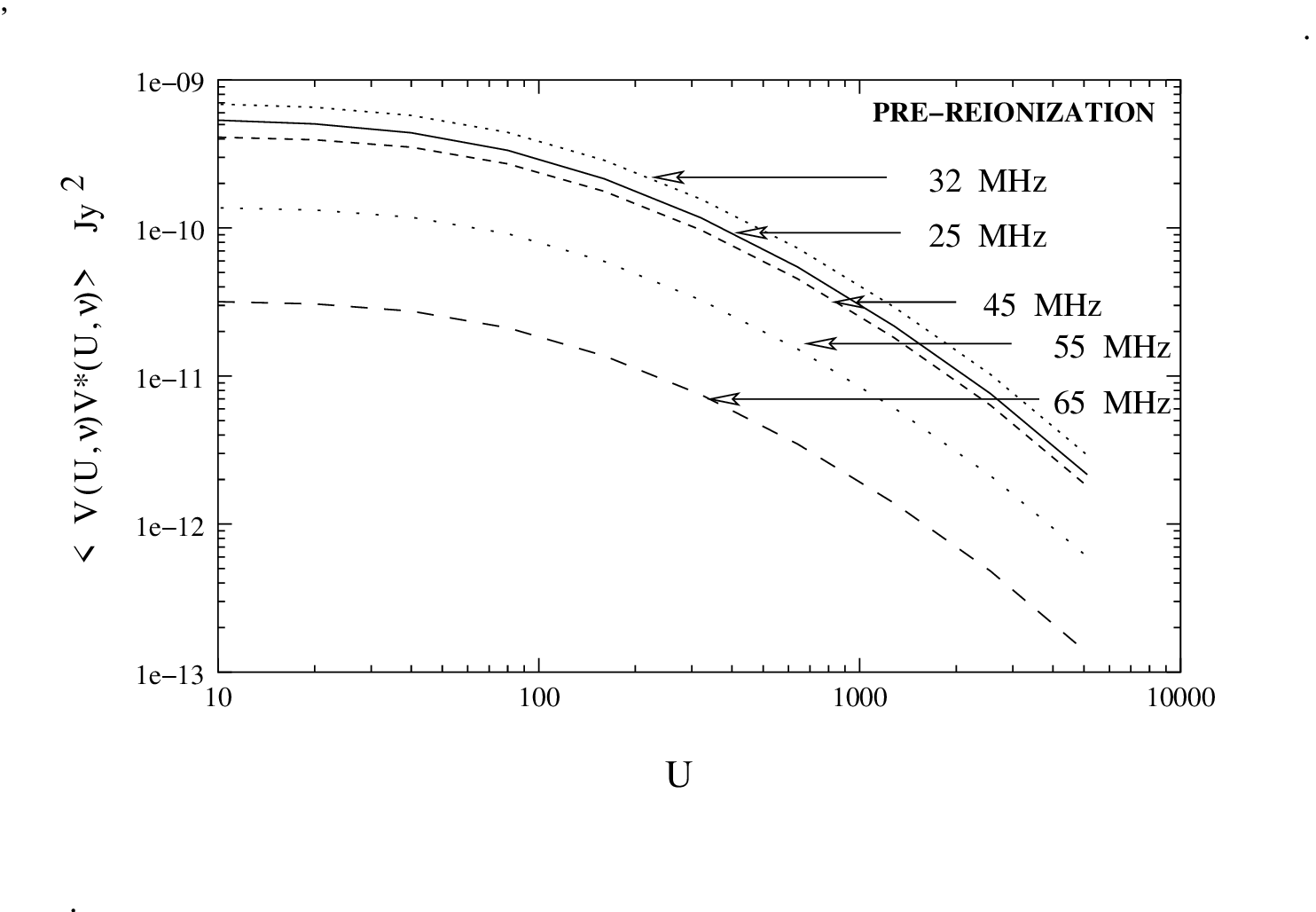}
\caption{This  shows how the visibility  correlation $\langle
  V(\u,\nu) V^{*}(\u,\nu) \rangle$   varies with baseline $U$. The
  results are shown for the different frequencies shown in the
  figure.} 
\label{fig:a4}
\end{figure}

The signal here will be absorption features in the CMBR spectrum. The
visibility correlations  record the angular fluctuations in this
decrement.  The results for $\Delta \nu=0$ are shown in Figure
\ref{fig:a4}.  The 
signal is maximum at a frequency $\sim 32 \, {\rm MHz}$ where the
perturbations in the HI density are most efficient in absorbing the
CMBR flux \citep{bharad5}. 
The signal strength at any fixed frequency does not vary with $U$ in
the range  of baselines $10$ to $100$, after which it falls rapidly,
falling by an order of magnitude by $U=1000$. Also, with varying
frequency the signal falls off quite fast  beyond $\sim 45 \,{\rm
  MHz}$. Frequencies  smaller than $25 \, {\rm MHz}$  will be severely 
affected by the ionosphere and have not been shown. The visibility
correlations as a function $U$ have a very similar behaviour at all the
frequencies, with just the amplitude of the signal changing with
frequency. The shape of these curves directly reflect the power
spectrum at the relevant Fourier modes. 

\begin{figure}
\includegraphics[width=84mm]{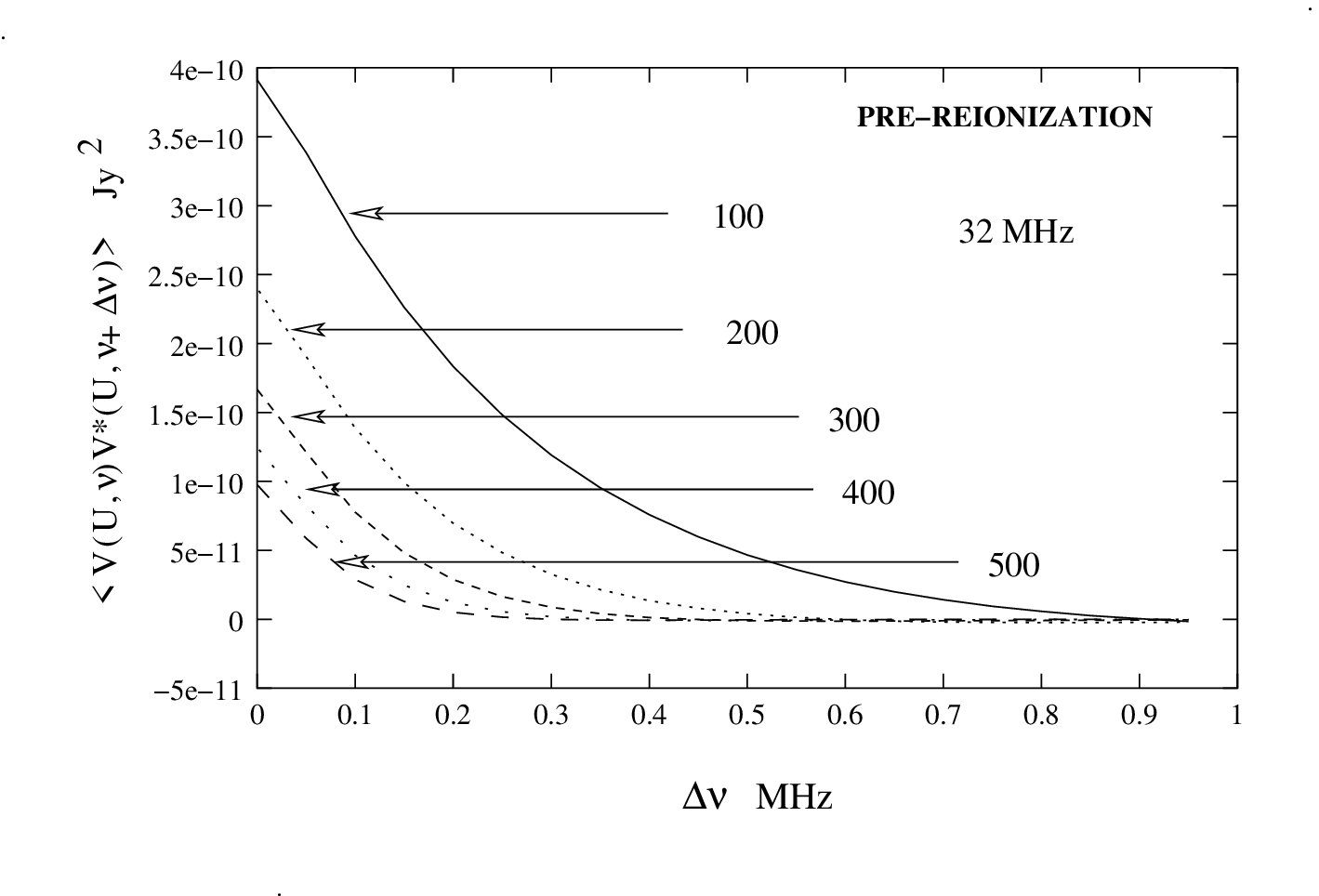}
\caption{This shows the correlation between the visibilities
  $V(U,\nu)$ and $V(U,\nu + \Delta\nu)$  expected for the same
  baseline $U$ at two slightly different frequencies. The result are
  shown for  different values of $U$ (shown in the figure) for  the $32
 {\rm MHz}$  band. }
\label{fig:a5}
\end{figure}

The visibility  correlation $\langle V(\u,\nu) V^{*}(\u,\nu+\Delta
\nu)$  for the same baseline and different frequency is shown for $32
\, {\rm MHz}$ in   Figure \ref{fig:a5}. The behaviour is similar at
other frequencies. The correlation falls rapidly as $\Delta \nu$ is
increased. The value of $\Delta \nu$ where the visibilities become 
uncorrelated scales as $\sim \rn/(\rnp U)$ and it is around $0.8 \,
{\rm MHz}$ for $U=100$.

\subsection{Reionization}

\begin{figure}
\includegraphics[width=84mm]{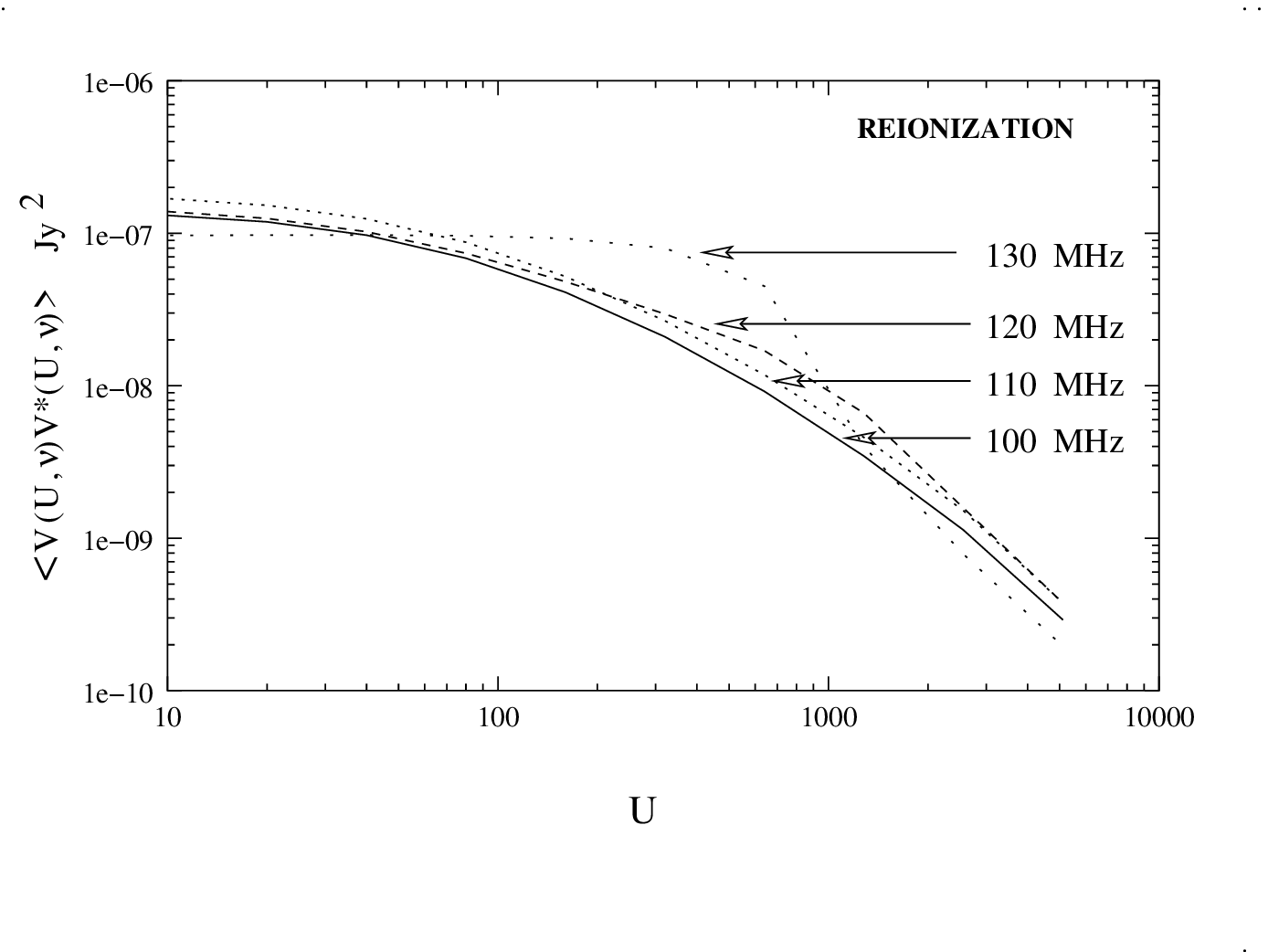}
\caption{This  shows how the visibility  correlation $\langle
  V(\u,\nu) V^{*}(\u,\nu) \rangle$   varies with baseline $U$. The
  results are shown for the different frequencies shown in the
  figure.} 
\label{fig:a6}
\end{figure}

The HI signal here is in emission. The results for  the visibility
correlation $\langle V(\u,\nu) V^{*}(\u,\nu) \rangle$ are shown as a
function of $U$ for different frequencies  in Figure \ref{fig:a6}.   At
$\nu \le 110 \, {\rm MHz}$ the hydrogen is largely neutral, and the
visibility correlations trace the dark matter power spectrum. The
ionization picks up at $\nu=120 \, {\rm MHz}$ (Figure \ref{fig:a2})
where ionized bubbles fill up $\sim 20 \%$ of the universe. The
centers of these bubbles are clustered with a bias $b_c=1.5$ with
respect to the underlying dark matter distribution. The presence of
these clustered bubbles reduces the signal at small baselines $U \le
400$. At $130 \, {\rm MHz}$ around $50 \%$ of the universe is occupied
by ionized bubbles. Here the clustering of the bubbles or of the
underlying dark matter is of no consequence, and the visibility
correlations are due to the discrete nature   (Poisson
distribution) of the ionized  bubbles. The visibility correlation
falls drastically for the large baselines which resolve out the
bubbles, and the signal at $U > 1000$ is due to the clustering of the
HI which follows the dark matter. The model for the reionization
breaks down when a large fraction of the universe is ionized, and we
do not present results beyond $130 \, {\rm MHz}$.  

An interesting feature when the visibility correlation is dominated by
the contribution from individual bubbles is that the fall in $\langle
V(\u,\nu) V^{*}(\u,\nu + \Delta \nu)$ with increasing $\Delta \nu$ is
decided by the ratio $R/\rnp$. This follows from the fact that the
power spectrum $P_{\HI}$ is now proportional to $W^2(k R)$ which falls
rapidly for $k \gg 1/R$, and the oscillating integral in
eq. (\ref{eq:a17}) cancels out when $\Delta \nu > R/ \rnp$. The value
of $\Delta \nu$ where the visibilities become uncorrelated depends
only on the size of the bubbles $R$  and does not change with $U$. 

\begin{figure}
\includegraphics[width=84mm]{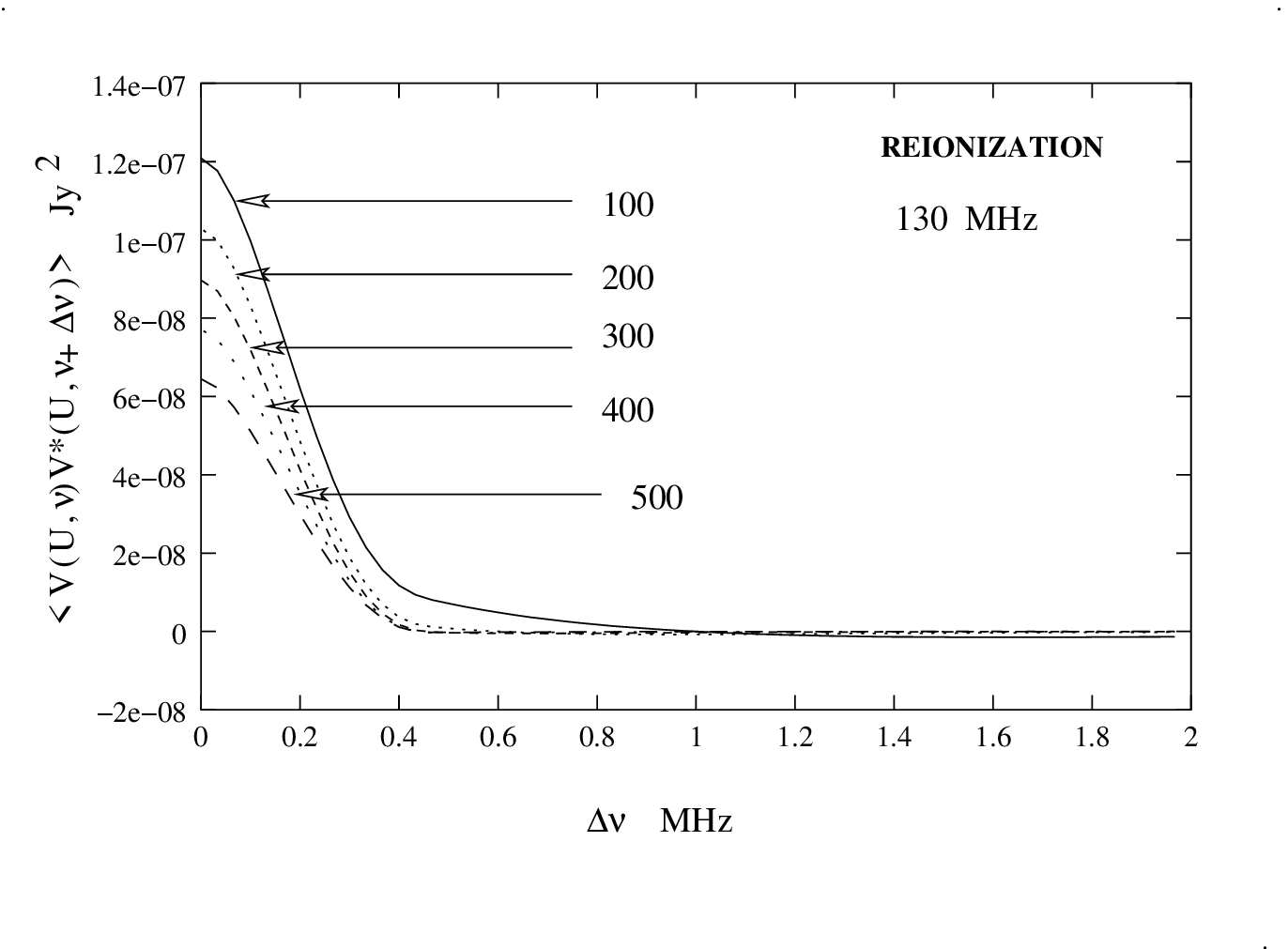}
\caption{This shows the correlation between the visibilities
  $V(U,\nu)$ and $V(U,\nu + \Delta\nu)$  expected for the same
  baseline $U$ at two slightly different frequencies. The result are
  shown for  different values of $U$ (shown in the figure) for  the $130
 {\rm MHz}$  band. }
\label{fig:a7}
\end{figure}

It should be noted that studies of the growth of  ionized  bubbles  
(eg. \citealt{furlanetto1}) show them to have a range of sizes at any
given  epoch during  reionization. This will possibly weaken some 
of the effects like the sharp drop in visibility correlations
(Figure \ref{fig:a6}) and the $\Delta \nu$  dependence (Figure
\ref{fig:a7})  calculated  here assuming bubbles of a fixed 
size.

\subsection{Post-reionization}
\begin{figure}
\includegraphics[width=84mm]{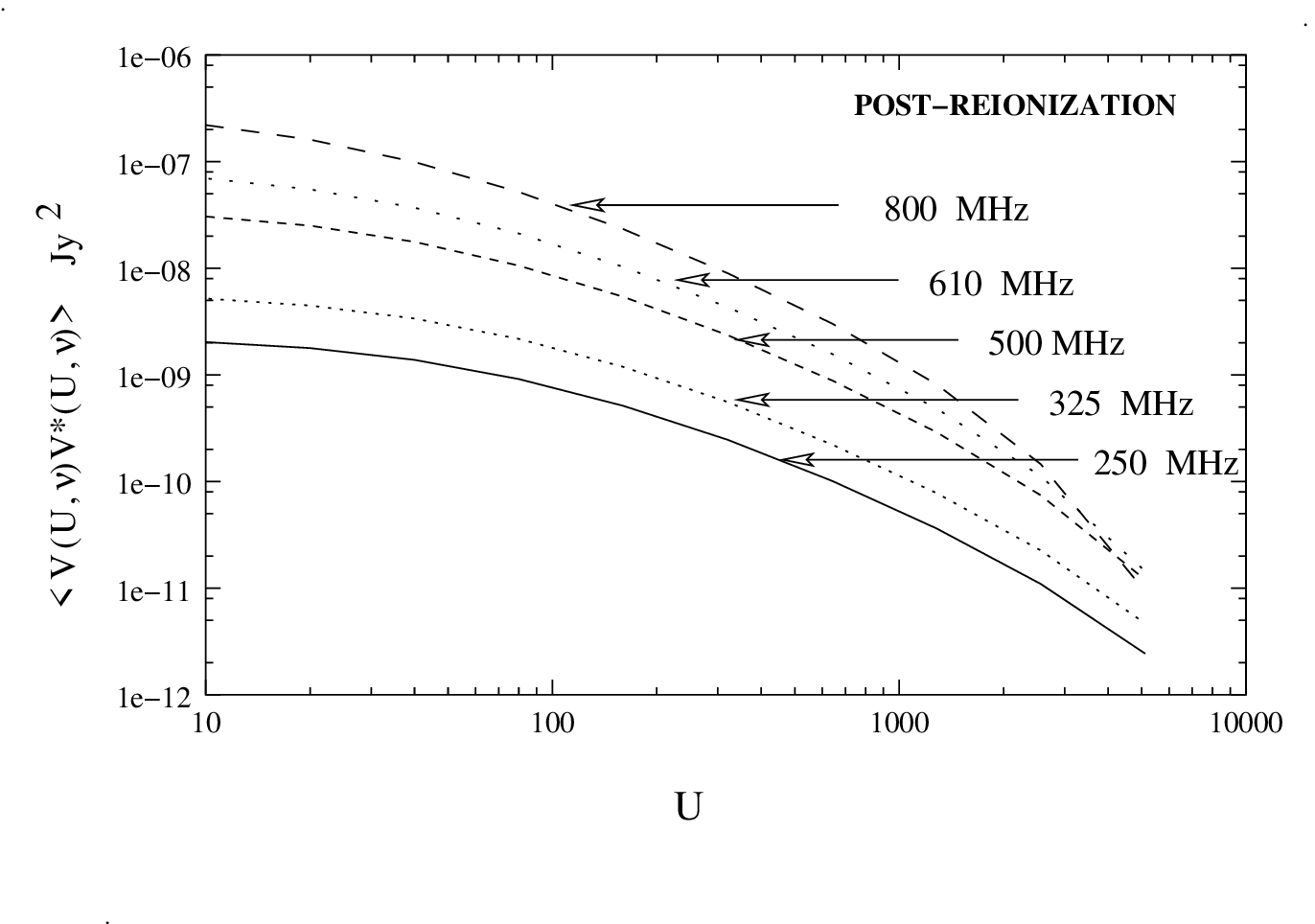}
\caption{This  shows how the visibility  correlation $\langle
  V(\u,\nu) V^{*}(\u,\nu) \rangle$   varies with baseline $U$. The
  results are shown for the different frequencies shown in the
  figure.} 
\label{fig:a8}
\end{figure}

\begin{figure}
\includegraphics[width=84mm]{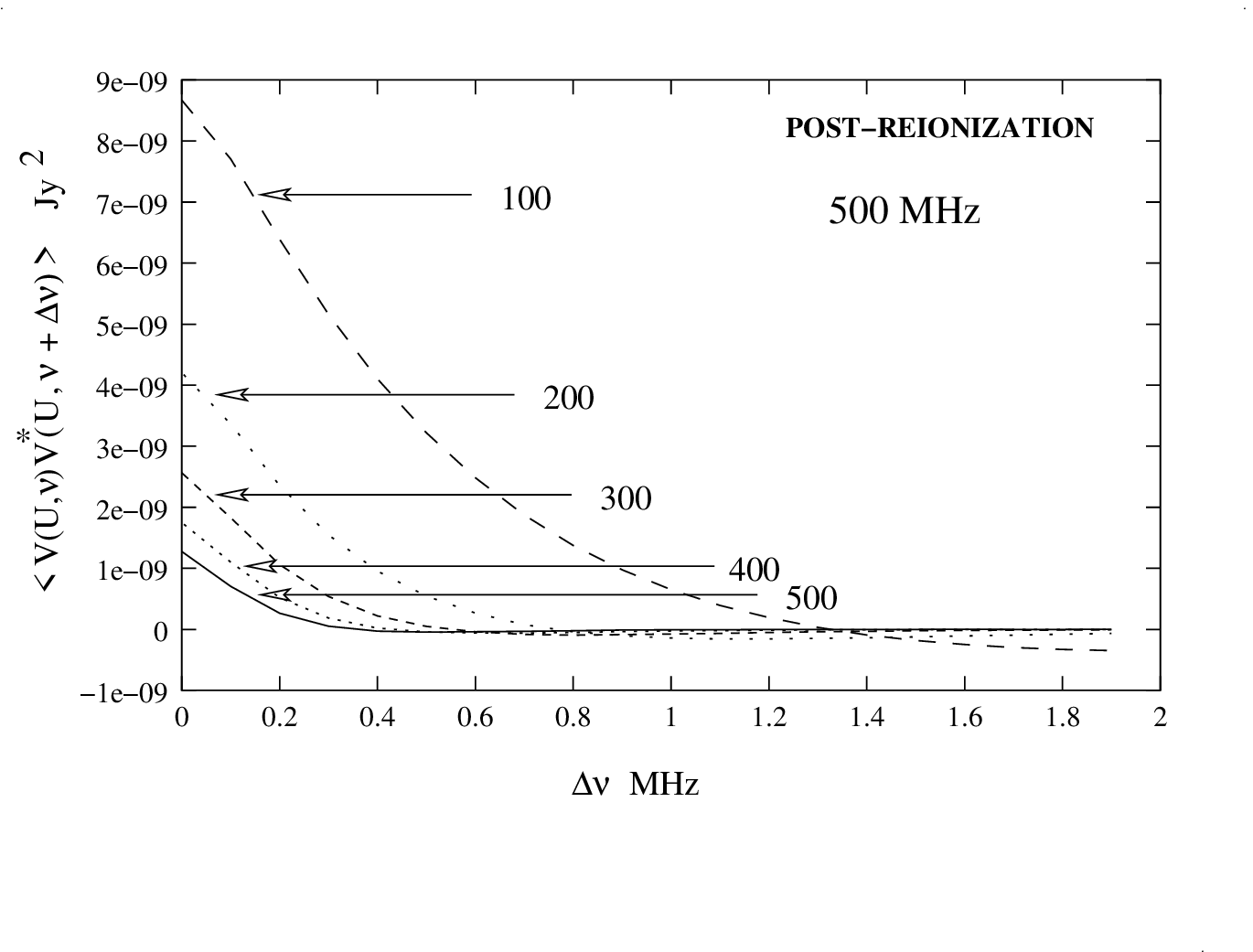}
\caption{This shows the correlation between the visibilities
  $V(U,\nu)$ and $V(U,\nu + \Delta\nu)$  expected for the same
  baseline $U$ at two slightly different frequencies. The result are
  shown for  different values of $U$ (shown in the figure) for  the $500
 {\rm MHz}$  band. }
\label{fig:a9}
\end{figure}

The predictions for $\langle V(\u,\nu) V^{*}(\u,\nu) \rangle$ are
shown as a function of $U$ for a variety of frequencies in Figure
\ref{fig:a8}. At all frequencies the correlations are nearly
independent of $U$ in the range $10$ to $100$, and then fall by
nearly an order of magnitude by $U=1000$. The HI in this era  has been
assumed to trace the dark matter, and the shape of the curve showing
the visibility correlation as a function of $U$  is decided by the 
dark matter power spectrum at the relevant Fourier modes. The
correlations increase with frequency,  a reflection of the fact that
the matter power spectrum grows with time.  It may be noted that the
signal also depends on $\Omega_{gas}$ which we have assumed to be a
constant over redshift, and the evolution of the signal 
probes both the growth of the power spectrum and the evolution of
$\Omega{gas}$. 
At a fixed baseline,  the
correlation $\langle V(\u,\nu) V^{*}(\u,\nu+\Delta \nu) \rangle$ falls
with increasing $\Delta \nu$. The value of $\Delta \nu$ where  the
visibilities become uncorrelated scales as  $\Delta \nu \sim \rn/\rnp
U$. The predictions for $\nu=500 \, {\rm   MHz}$ are shown in Figure
\ref{fig:a9} where the visibility at $U=100$ becomes uncorrelated at
$\Delta \nu \sim 1.3 \, {\rm   MHz}$. The behaviour is similar at the
at the other frequencies.

\section{Summary and Discussion.} 
We have proposed that  correlations between the visibilities measured 
at different baselines and frequencies in radio-interferometric
observations be used to quantify and interpret the redshifted 21 cm
radiation from cosmological HI.  This statistics has an inherent
advantage over other quantities which have been proposed for the same
purpose in that it deals directly with visibilities which are the
primary quantities measured in radio-interferometric
observations. There is a further advantage in that the system noise in
the visibilities measured  at different baselines and frequencies is
uncorrelated.  

The visibility correlations, we show, have a very simple relation to
the power spectrum (equation \ref{eq:a17}), the latter being the
statistical quantity most  commonly used to quantify the cosmological
matter distribution. Also, there is a direct relation between the
dimensionless baseline $U$ and the Fourier mode probed by the
visibilities at that particular baseline. 

We have made predictions for the signal  expected  over the entire
evolution of the HI starting from the dark ages to the present epoch. 
The magnitude of the signal depends on $\theta_0 \approx 0.6 \times 
\theta_{FWHM}$ the size of the beam  of the individual antenna elements
in the array. Our predictions are  for  $\theta_0=1^{\circ}$ at
$325 {\rm MHz}$ which is appropriate for the GMRT.  At other
frequencies we have used  $\theta_0 \propto \lambda$  in making our
predictions.  The results presented here can be easily used to make 
predictions for other telescopes using the fact that the visibility
correlations scale as $\theta_0^2$. The shape of the curves showing
$\langle V(\u,\nu) V^{*}(\u,\nu + \Delta \nu) \rangle $ as functions
of $U$ or $\Delta \nu$ have no dependence on $\theta_0$ or any other
parameter of the individual antennas.  Finally, a reminder that our
analysis assumes $\theta_0 \ll 1$ where the curvature of
the sky may be neglected. The analysis will be more complicated for
an array of small antennas where the individual elements have a very
wide beam, but possibly the qualitative features of this analysis will
hold.   

One of the salient features of our analysis is that for all the eras
the visibility correlation signal from redshifted HI is maximum at
small baselines. In the situation where the signal arises mainly from
the large scale clustering of the HI whose distribution follows the
dark matter   the signal is nearly constant up to $U \sim 100$ 
after which the signal falls rapidly dropping by an order of magnitude
by $U \sim 1000$. In the situation when the process of reionization is
underway and the HI distribution is patchy, the behaviour of the
visibility correlation is decided by the size of the ionized
regions. In our model for reionization where there are spherical
ionized bubbles of comoving radius $R=3 \,h^{-1} {\rm Mpc}$, the
visibility correlation is a constant to $U \sim 400$, the larger
baselines are not sensitive to the ionized spheres.   Details of the
reionization model aside, the baseline dependence of the visibility
correlations is sensitive to size  of the ionized patches and it
should be possible to extract this information. Also, an optimal
observational strategy would preferentially sample the small baselines
where the signal is largest. 

We next shift  our attention to the correlation between  the
visibilities at the same baseline but at two different frequencies
with separation $\Delta \nu$. We see that the correlations fall off
very fast as $\Delta \nu$ is increased. In the situation where the
visibility correlation signal is from the large scale gravitational
clustering of the HI the 
value of $\Delta \nu$ where the visibilities become uncorrelated
scales as $\rn/(\rnp U)$ and it is around $1 \, {\rm MHz}$ at $U=100$
at all frequencies. In the situation where the signal is from ionized
patches, the value of $\Delta \nu$ where the visibilities become
uncorrelated is decided by $R/\rnp$ and it is around $0.4 \, {\rm
  MHz}$ independent of the value of $U$. This feature should allow us
to determine from the observations whether it is the large scale
gravitational clustering of the HI or the ionized patches which
dominates  the visibility correlation. The latter
is expected to dominate during  the epoch when reionization is
underway. It may be noted that the value $0.4 \, {\rm MHz}$ quoted
here is specific to the model of reionization which we have adopted. 
Though the value is model dependent, we expect that the qualitative 
features discussed here should continue to hold even in a more general
situation.

The main challenge in observing cosmological HI is to  extract the HI
signal from the various contaminants which are expected to swamp this 
signal. The contaminants include  Galactic synchrotron
emission, free-free  emission from ionizing halos  \citep{oh}, faint  
radio loud quasars \citep{dimat1} and synchrotron emission from low
redshift galaxy clusters \citep{dimat}. Fortunately, all of these
foregrounds have smooth continuum spectra whose contribution varies
slowly with frequency. The contribution from these contaminants to the
visibilities at the same baseline but two frequencies differing by
$\Delta \nu$ will remain correlated for large values of $\Delta \nu$,
whereas the contribution from the HI signal will be uncorrelated
beyond something like $1 \, {\rm MHz}$ or even less depending on the
value of $U$. It is, in principle, straightforward to fit the
visibility correlation at large $\Delta \nu$ and remove any slowly
varying component thereby separating the contaminants from the HI
signal.

Finally, we briefly discuss the level of visibility correlation
detectable in a radio-interferometric observation.  We consider
an array of $N$ antennas, the observations lasting a time duration
$t$, with frequency channels of 
width $\delta \nu$ spanning a total bandwidth $B$.  It should be noted
that the effect of a finite channel width $\delta \nu$ has not been
included in our calculation which assumes infinite frequency
resolution. This effect can be easily included by convolving our
results for the visibility correlation with the frequency
response function of a single channel. Preferably, $\delta \nu$
should be much smaller than the frequency separation at which the
visibility correlation become  uncorrelated.  We use $S$ to denote the  
frequency separation within which the visibilities are correlated,  and
beyond which they become uncorrelated. The rms. noise in the a single 
visibility correlation is \citep{thomp}
\begin{equation}
\sqrt{ \langle (V V*)^2 \rangle} = \left( \frac{2 k_B T_{SYS}}{A_{ef}}
  \right)^2 \frac{1}{\delta \nu \, t}
\label{eq:di1}
\end{equation}
where $T_{SYS}$ is the system temperature and $A_{ef}$ is the effective
area of a single antenna. The noise contribution  will be reduced by a
factor $1/\sqrt{N_o}$ if we  combine $N_o$ independent samples of the 
visibility correlation.  A possible observational strategy for  a
preliminary detection of the HI signal would be to combine the
visibility correlations at all baselines and frequency separations
where there is a reasonable amount of signal. This gives
$N_o=[N(N-1)/2] \,  (B/\delta \nu) \, (S/\delta \nu)$ whereby 
\begin{equation}
\sqrt{\langle (V V*)^2 \rangle }= \left( \frac{2 k_B T_{SYS}}{A_{ef}}
  \right)^2  \, \frac{1}{t} \, \sqrt{\frac{2}{N (N-1) \, B \, S}}
\label{eq:di2} \,.
\end{equation}

Using values for the GMRT (at $ \sim 325\, {\rm MHz}$), $(2 k_B
T_{SYS}/A_{ef})=144 \, {\rm Jy}$,  $B=16 \, {\rm MHz}$, there are $14$ 
antennas within $U \le 1000$  where the signal is 
strong and $S=0.5 \,  {\rm MHz}$ beyond which the signal is
uncorrelated we find that it is possible to achieve noise levels of 
$\sqrt{\langle (V V*)^2 \rangle } = 1.0 \times 10^{-9} \, {\rm Jy}^2$
which is below the signal  with  $200 \, {\rm hrs}$ of integration. 

 We shall investigate in detail issues related to detecting the
HI signal, the necessary integration time for different telescopes
and signal extraction in a forthcoming paper. 

\section*{Acknowledgments}
SB would like to thank  Jayaram  Chengalur for useful discussions. 
 SB would also like to acknowledge BRNS, DAE, Govt. of India,for
 financial support through sanction No. 2002/37/25/BRNS.   
SSA  is supported by a junior research fellowship of the
Council of Scientific and Industrial Research (CSIR), India.
\appendix 
\section{The optical depth of the redshifted HI 21 cm line.}
 The optical depth is defined as
\begin{equation}
  \tau_\nu(l) = \int^l_{l0}\alpha_\nu(l^{\prime})dl^\prime 
\end{equation} 
where $\alpha_\nu$ is the absorption coefficient for the redshifted 21 cm
line given by
\begin{equation}
\alpha_\nu=\frac{h_P \nue}{4\pi}\phi(\nu)[n_0 \Bo-n_1 \Bon] \,.
\end{equation}
Here $\phi(\nu)$ is the line profile function, and $B_{01}$ and
$B_{10}$ are the Einstein coefficients  \citep{rybicki} for the 21 cm
line.  
We use $r(\nu)$ to denote the comoving distance from which the $1420
  \,   {\rm MHz}$ radiation will be redshifted to the frequency $\nu$
  for 
  an observer at present.  The HI  radiation from a physical line element
  $\Delta  l$  at the epoch when the HI radiation originated  will be
  spread  over a   frequency range  $\Delta  \nu=(\partial
  r(\nu)/\partial 
  \nu)^{-1} \,  (\Delta   l/a)$  at the observer. Here $a=1/(1+z)$ is
  the scale factor. 
 The line profile function $\phi(\nu)$,  defined at the epoch when the
  HI radiation originated,  is a  step function of width
  $a/\Delta \nu$. It then follows that 
\begin{equation}
\int \phi(\nu) \, d l =a^2 \frac{\partial r(\nu)}{\partial
  \nu}
\end{equation}
and the optical  depth can be written as  
\begin{equation}
\tau_\nu = \frac{h_P \nue}{4\pi} n_0 \Bo(1-\frac{g_0}{g_1}
\frac{n_0}{n_1}) a^2 \mid \frac{\partial  r(\nu)}{\partial\nu} \mid
\,. 
\end{equation}
The Einstein coefficients are related by
$\Bo = 3\Bon =\frac{3\lambda_e^3}{2\hp c}A_{10}$ , where
$A_{10}=2.85\times10^-{15}$  $s^{-1} $ is the spontaneous emission
coefficient of the 21 cm line . So the optical depth finally becomes
\begin{equation}
\tau_{\nu}=\frac{ 3 \nH \hp c^2  A_{10} a^2}{32 \pi \kb \Ts \nue}
 \mid \frac{\partial r(\nu)}{\partial\nu}  \mid  
\end{equation}
where $\nH$ is the number density of HI atoms. 
In presence of peculiar velocity $r(\nu)$ is given by
\begin{equation}
r(\nu)=\int^{1}_{\frac{\nu}{\nue(1-v/c)}}\hspace{.2cm}\frac{c
  \,da}{a^2\,H(a)} 
\end{equation}
where $v$ is the line of sight component of the peculiar
velocity of the HI. This  gives us 
\begin{equation}
\mid\frac{\partial r(\nu)}{\partial \nu}\mid=\frac{\lambda_e}{a^2
  H(z)}[1- \frac{1}{a H(z)}\frac{\partial v}{\partial \rn}]
\end{equation}
where $\rn$ is the same as $r(\nu)$ without the effect of peculiar
velocities, and where we have retained only the most dominant term
in the peculiar velocity. 
Using this it is possible to write the optical depth as 
\begin{eqnarray}
\tau_{\nu} &=& \frac{4.0 \, \mK }{\Ts}\, \left(\frac{\Omega_b
  h^2}{0.02}\right) 
\left( \frac{0.7}{h} \right) \frac{H_0}{H(z)} (1+z)^3 \,  \nonumber \\ 
&\times&   \frac{\rho_{\HI}}{\bar{\rho}_{\H}} \,   \left[1-\frac{(1+z)}{ 
  H(z)}\frac{\partial v}{\partial \rn}\right]    
\end{eqnarray}
where $\rho_{\HI}/ \bar{\rho}_{\H}$ is the ratio 
of the neutral  hydrogen to the mean hydrogen density.


\begin{thebibliography}{99}
\bibitem [\protect\citeauthoryear{Bagla, Nath and
 Padmanabhan}{1997}]{bagla1} 
 Bagla J.S., Nath B. and Padmanabhan T. 1997, MNRAS 289, 671
\bibitem [\protect\citeauthoryear{Bagla \& White}{2002}]{bagla2} 
Bagla J.S. and White M. 2002, astro-ph/0212228
\bibitem [\protect\citeauthoryear{Barkana \& Loeb}{2001}]{barkana}
  Barkana  R. \& Loeb A.,2001,Phys.Rep.,349,125
\bibitem [\protect\citeauthoryear{Becker et al.}{2001}]{becker}
  Becker,R.H.,et al.,2001,AJ,122,2850
\bibitem [\protect\citeauthoryear{Bharadwaj, Nath \&
    Sethi}{2001}]{bharad1} 
  Bharadwaj S., Nath B. \& Sethi S.K. 2001, JApA.22,21
\bibitem [\protect\citeauthoryear{Bharadwaj \&  Sethi}{2001}]{bharad2} 
Bharadwaj, S.~\&  Sethi, S.~K.\ 2001, JApA, 22, 293  
\bibitem [\protect\citeauthoryear{Bharadwaj \&  Pandey}{2003}]{bharad3} 
Bharadwaj, S.~\&  Pandey, S.~K.\ 2003, JApA, 24, 23  
\bibitem [\protect\citeauthoryear{Bharadwaj \& Srikant}{2004}]{bharad4} 
Bharadwaj, S. ~\&  Srikant p,s. 2004, JApA, in press
\bibitem [\protect\citeauthoryear{Bharadwaj \& Ali}{2004}]{bharad5} 
Bharadwaj S. \& Ali S. S. 2004, \mnras,in Press, 
\bibitem [\protect\citeauthoryear{Chen \&   Miralda-Escude}{2004}]{chen}
Chen, X. \&   Miralda-Escude,J.,2004, \apj, 602, 1
\bibitem [\protect\citeauthoryear{Ciardi \& Madau}{2003}]{ciardi}
Ciardi,B.\& Madau,P. 2003,\apj,596,1
\bibitem [\protect\citeauthoryear{DiMatteo. et al.}{2004}]{dimat}
  DiMatteo,T.,Ciardi,B.,Miniati,F.2004,\mnras,submitted,astro-ph/0402332
\bibitem [\protect\citeauthoryear{DiMatteo. et al.}{2002}]{dimat1}
  DiMatteo,T.,Perna R.,Abel,T.,Rees,M.J.2002,\apj,564,576
\bibitem [\protect\citeauthoryear{Fan et al.}{2002}]{fan}
  Fan,X.,et al. 2002,AJ,123,1247
\bibitem [\protect\citeauthoryear{Furlanetto,Sokasian. \&
  Hernquist}{2003}]{furlanetto} 
  Furlanetto, S. R.,Sokasian  A. \& Hcernquist
  L.2003,astro-ph/0305065 
\bibitem [\protect\citeauthoryear{Furlanetto, Zaldarriaga \&
  Hernquist}{2004}]{furlanetto1}  
  Furlanetto, S. R., Zaldarriaga, M. \& Hcernquist
  L. 2004, astro-ph/0403697
\bibitem [\protect\citeauthoryear{Gnedin \&   Ostriker}{1997}]{gnedin}  
  Gnedin, N. Y. \&  Ostriker, J. P,1997,\apj,486,581
\bibitem [\protect\citeauthoryear{Gnedin \& Shaver}{2003}]{gnedin2}
  Gnedin, N. Y. \& Shaver, P. A.2003,astro-ph/0312005
\bibitem [\protect\citeauthoryear{Gruzinov \& Hu}{1998}]{gruz}
Gruzinov, A. \& Hu, W. 1998,\apj,508,435
\bibitem [\protect\citeauthoryear{Hogan \& Rees}{1979}]{hogan}
Hogan, C. J. \& Rees, M. J., 1979, MNRAS,188,791
\bibitem [\protect\citeauthoryear{ Hui \& Haiman}{2003}]{hui}
  Hui, L., \& Haiman, Z., 2003, ApJ,596,9
\bibitem [\protect\citeauthoryear{Iliev et al.}{2002}]{isfm}
  Iliev,I.T.,Shapiro,P.R.,Farrara,A.,Martel,H.2002,\apj,572,L123
\bibitem [\protect\citeauthoryear{Iliev et al.}{2003}]{iliev}
  Iliev,I.T.,Scannapieco,E.,Martel,H.,Shapiro,P.R. 2003,\mnras,341,81
\bibitem[\protect\citeauthoryear{Kaiser}{1987}]{kais}
  Kaiser N. 1987, \mnras,227,1
\bibitem[\protect\citeauthoryear{Kumar, Padmanabhan \&
    Subramanian}{1985}]{kumar} 
 Kumar A., Padmanabhan T. and Subramanian K., 1995, MNRAS, 272,
      544
\bibitem [\protect\citeauthoryear{Knox at al.}{1998}]{knox}
  Knox L.,Scoccimarro  R. \& Dodelson  S.1998,Physical Review Letters,81,2004
\bibitem [\protect\citeauthoryear{Lahav et al.}{1991}]{lahav}
Lahav, O., Lipje, P. B., Primack, J. R. and Rees, M. J. 1991,
MNRAS,251, 128
\bibitem [\protect\citeauthoryear{Lanzetta, Wolfe \&
    Turnshek}{1998}]{lanzetta} 
  Lanzetta, K. M., Wolfe, A. M., Turnshek, D. A. 1995,
    ApJ, 430, 435
\bibitem [\protect\citeauthoryear{Loeb \& Zaldarriaga}{2003}]{lz}
  Loeb  A. \& Zaldarriaga,  M.,2003,astro-ph/0312134
\bibitem [\protect\citeauthoryear{Madau, Meiksin \& Rees}{1997}]{madau}
  Madau  P., Meiksin  A. \& Rees, M. J.,1997,\apj,475,429
\bibitem [\protect\citeauthoryear{Miralda-Escude}{2003}]{miralda}
  Miralda-Escude,J.,2003,Science 300,1904-1909
\bibitem [\protect\citeauthoryear{Morales}{2004}]{morales1}
Morales, M. F.  2004, preprint, astro-ph/0406662
\bibitem [\protect\citeauthoryear{Morales \& Hewitt}{2003}]{morales}
Morales, M. F. and Hewitt, J., 2003, ApJ, Submitted (astro-ph/0312437) 
\bibitem [\protect\citeauthoryear{Oh \& Mack}{2003}]{oh}
  Oh,S.P.,\& Mack,K.J.,2003,\mnras,346,871
\bibitem [\protect\citeauthoryear{P\'eroux et al.}{2001}]{peroux}
P\'eroux, C., McMahon, R. G., Storrie-Lombardi,
    L. J. \& Irwin, M .J. 2003, MNRAS, 346, 1103 
  \bibitem [\protect\citeauthoryear{Rybicki. \& Lightman}{1979}]{rybicki}
  Rybicki, G. B. \&  Lightman, A. P.,1979,Radiative Processes in
  Astrophysics.Willey,New York,pp. 29-32
\bibitem [\protect\citeauthoryear{Santos, Cooray \& Knox}{2004}]{santos}
Santos, M.G., Cooray, A. \& Knox, L. 2004, preprint,  astro-ph/0408515 
\bibitem [\protect\citeauthoryear{Saini, Bharadwaj \&
    Sethi}{2001}]{saini} 
    Saini T., Bharadwaj  S. \& Sethi, K. S. 2001, ApJ, 557, 421
  \bibitem [\protect\citeauthoryear{Scott \& Rees}{1990}]{scott}
  Scott  D. \& Rees, M. J., 1990, \mnras,247,510
\bibitem [\protect\citeauthoryear{Shaver et al.}{1999}]{shaver}
  Shaver, P. A., Windhorst, R, A., Madau, P.\& de Bruyn,
  A. G.,1999,Astron. \& Astrophys.,345,380 
\bibitem [\protect\citeauthoryear{Spergel et al.}{2003}]{spergel}
  Spergel, D. N.,et al. 2003,ApJS,148,175
\bibitem [\protect\citeauthoryear{Subramanian \&
    Padmanabhan}{1993}]{subramanian} 
Subramanian K. and Padmanabhan T., 1993, MNRAS, 265, 101
\bibitem [\protect\citeauthoryear{Sunyaev and Zeldovich}{1972}]{suny}  
  Sunyaev, R. A. \& Zeldovich, Ya. B., 1972,Astron. \& Astrophys,22,189
\bibitem [\protect\citeauthoryear{Storrie-Lombardi, McMahon \& Irwin}
    {1996}]{lombardi} 
 Storrie--Lombardi, L. J., McMahon, R. G., Irwin, M. J. 1996, MNRAS,
   283, L79
\bibitem [\protect\citeauthoryear{Tegmark et al.}{2003a}]{tegmarka}
Tegmark, M., et al., 2003a, \apj, Submitted   (astro-ph/0310725) 
  %%%%%%
\bibitem [\protect\citeauthoryear{Tegmark et al.}{2003b}]{tegmarkb}
Tegmark, M., et al., 2003b, \apj, Submitted   (astro-ph/0310723) 
\bibitem [\protect\citeauthoryear{Theuns et al.}{2002}]{theuns}
Theuns, T. et al., 2002, ApJ,567,L103
\bibitem [\protect\citeauthoryear{Thompson, Moran \& Swenson}{1986}]{thomp}
Thompson, A.R., Moran, J.M. and Swenson, G.W., Jr. 1986,
Interferometry and Synthesis in Radio Astronomy, 
John Wiley and Sons, New York, pp 162-165
\bibitem [\protect\citeauthoryear{Tozzi et al.}{2000}]{tozzi}
 Tozzi.P.,Madau.P.,Meiksin. A.,Rees,M.J.,2000,\apj,528,597
\bibitem [\protect\citeauthoryear{Zaldarriaga, Furlanetto \&
    Hernquist}{2003}]{zald} 
Zaldarriaga M., Furlanetto, S. R. \& Hernquits L.,  2003, \apj,
    SUbmittedy
\end{thebibliography}
\end{document}